%% file: main.tex
\documentclass[10pt,a4paper,twocolumn]{article}
\usepackage[margin=1.8cm]{geometry}
\usepackage[T1]{fontenc}
\usepackage{microtype}
\usepackage[english]{babel}
\usepackage{authblk}
\usepackage{xcolor}
\usepackage{hyperref}
\usepackage{orcidlink}

\usepackage[numbers]{natbib}
\usepackage{verbatim, amsmath, amsfonts, amssymb, amsthm}
\usepackage{graphicx}
\usepackage{caption} 
\usepackage{subfigure}
\usepackage{epstopdf}
\usepackage{booktabs}
\usepackage{moreverb,url}
\usepackage{soul}
\usepackage{siunitx}
\usepackage{acronym}
\usepackage{float}
\usepackage{graphicx,url,algorithm}
\usepackage{algpseudocode}
\usepackage{pifont}%
\usepackage{dirtytalk}
\usepackage{tabularx}
\usepackage{cleveref}
\usepackage{color}
\usepackage{tikz}
\usepackage[normalem]{ulem}
\usepackage{acronym}
\usepackage{placeins}
\usepackage{multirow}

\newcommand\BibTeX{{\rmfamily B\kern-.05em \textsc{i\kern-.025em b}\kern-.08em
T\kern-.1667em\lower.7ex\hbox{E}\kern-.125emX}}

\usepackage{tabularx}

\newcolumntype{N}{>{\raggedright\arraybackslash}p{0.08\columnwidth}}
\newcolumntype{U}{>{\raggedright\arraybackslash}p{0.2\columnwidth}}
\newcolumntype{D}{X}
\newcolumntype{H}{>{\raggedright\arraybackslash}X}
\newcolumntype{M}{>{\centering\arraybackslash}X}
\hypersetup{
    colorlinks=true,
    linkcolor=black,
    citecolor=black,
    urlcolor=black
}

\acrodef{MPC}{Model Predictive Control}
\acrodef{RMSE}{Root Mean Square Error}
\acrodef{IOB}{Insulin on Board}
\acrodef{BG}{blood glucose}
\acrodef{APS}{Artificial Pancreas Systems}
\acrodef{AID}{Automated Insulin Delivery}
\acrodef{SAP}{Sensor-Augmented Pump}
\acrodef{AP}{Artificial Pancreas}
\acrodef{CLC}{Closed-Loop Control}
\acrodef{T1D}{Type 1 Diabetes}
\acrodef{HbA1c}{Hemoglobin A1c}
\acrodef{TDI}{Total Daily Insulin}
\acrodef{IIT}{Intensive Insulin Therapy}
\acrodef{CR}{carbohydrate-to-insulin ratio}
\acrodef{CF}{correction factor}
\acrodef{CGM}{Continuous Glucose Monitoring}
\acrodef{TIR}{Time-In-range}
\acrodef{CVGA}{control variability grid analysis}
\acrodef{FDA}{Food and Drug Administration}
\acrodef{LBGI}{low blood glucose index}
\acrodef{HBGI}{high blood glucose index}
\acrodef{DIA}{duration of insulin action}
\acrodef{SMBG}{self-monitoring blood glucose}
\acrodef{CHO}{carbohydrate}
\acrodef{IIR}{insulin infusion rate}
\acrodef{PK}{pharmacokinetic}
\acrodef{EGP}{endogenous glucose production}
\acrodef{IG}{interstitial glucose}
\acrodef{SD}{standard deviation}
\acrodef{PDF}{probability density function}
\acrodef{GIR}{glucose infusion rate}
\acrodef{MSE}{mean squared error}
\acrodef{NE}{net effect}
\acrodef{IVS}{insulin variability signal}
\acrodef{LTI}{linear time invariant}
\acrodef{DSS}{decision support systems}
\acrodef{MDI}{Multiple Daily Injections}
\acrodef{PH}{Prediction Horizon}
\acrodef{GRM}{Glucoregulatory Model}
\acrodef{T1DEXI}{Type 1 Diabetes Exercise Initiative}
\acrodef{SA}{Sensitivity Analysis}
\acrodef{HGM}{Hovorka Glucoregulatory Model}
\acrodef{DT}{Digital Twin}
\acrodef{LES}{Lowest Error Strategy}
\acrodef{EKF}{Extended Kalman Filter}

\usepackage{fancyhdr}

\fancypagestyle{firstpage}{
	\fancyhf{}
	\fancyfoot[C]{\scriptsize\textit{This work has been submitted to IFAC for possible publication.}}
	\fancyfoot[R]{\thepage}

}

\title{
\textbf{Sensitivity-driven Personalization of a Glucoregulatory Model
for Digital Twin Therapeutics in Type 1 Diabetes}
}

\author[1,2,3]{
Clara Escorihuela-Altaba\,\orcidlink{0009-0003-1680-920X}
}

\author[1,2]{
Vihangkumar V. Naik\,\orcidlink{0000-0002-2214-8052}
}

\author[1,2]{
Eleonora Manzoni\,\orcidlink{0000-0002-2445-6104}
}

\author[1,2]{
Jose Garcia-Tirado\,\orcidlink{0000-0002-9970-2162}\thanks{
Corresponding author: \href{mailto:jose.garcia@unibe.ch}
{jose.garcia@unibe.ch}
}
}

\affil[1]{
Department of Diabetes, Endocrinology, Nutritional Medicine, and Metabolism,
Inselspital, Bern University Hospital and University of Bern,
Bern, Switzerland
}

\affil[2]{
Diabetes Center Berne, Bern, Switzerland
}

\affil[3]{
Graduate School for Cellular and Biomedical Sciences,
University of Bern, Bern, Switzerland
}

\date{}

\begin{document}

\maketitle
\thispagestyle{firstpage}

\vspace{0.5cm}

\begin{abstract}
Digital twin technologies are increasingly applied in diabetes research to replicate individual metabolic behavior, but their reliability depends on accurate identification of parameters within glucoregulatory models. Traditional sensitivity analysis approaches often overlook the full range of dynamic inputs and intersubject variability, possibly causing influential parameters, especially those governing transient responses, to be missed. This study addresses these limitations by quantifying both magnitude and timing of parameter influence under dynamic input-output scenarios for nonlinear glucoregulatory models, and by examining whether a universal parameter ranking can be established across individuals.\\
Using the Hovorka model as a representative framework, parameter importance was assessed through a time‑series extension of Sobol sensitivity analysis. Sensitivity rankings were computed under four input conditions (full day profiles, isolated meal disturbances, insulin bolus injections, and postprandial responses), capturing the breadth of real‑world metabolic variability. These condition‑specific rankings were consolidated into a global metric. The resulting ranking was then used for participant‑specific parameter identification, enabling evaluation of how restricting identification to the most influential parameters affects predictive capabilities.
The study analyzed data from 192 individuals in the T1DEXI dataset. Across participants, parameters governing insulin appearance and clearance consistently exhibited strong influence. The analysis also revealed how parameter relevance shifts over time depending on input dynamics. Constraining identification to the sensitivity‑derived subset produced a 60\% reduction in Root Mean Square Error over three days of test data under both intervention-informed and forecasting modes, corresponding to scenarios where future meal and insulin inputs are assumed to be known or unknown within the prediction horizon, respectively.\\
The findings support the creation of a global parameter ranking that accommodates intersubject variability and dynamic inputs, reducing computational burden and accelerating parameter identification for digital twin applications.
\end{abstract}


\input{Sections/Introduction}

\input{Sections/Methods}

\input{Sections/Results}
\input{Sections/Discussion}

\input{Sections/Conclusion}

\input{Sections/declarations}

\appendix
\input{Sections/Appendix}

\bibliographystyle{elsarticle-num}

\bibliography{EDTbib}

\end{document}

%% file: Sections/Introduction.tex
\section{Introduction}
\label{sec:introduction}

\ac{AID} systems are now widely regarded as the standard of care for individuals with \ac{T1D} \cite{american2026}. In a closed-loop setup, the control algorithm calculates insulin delivery directly from \ac{CGM} sensor data and administers it via a subcutaneous pump. Current systems have shown reliable daytime glucose regulation \cite{gera2025}, improved nocturnal safety \cite{gera2025}, and a reduction in hypoglycemic events \cite{sherr2013}, and psychosocial benefits \cite{carlson2022,berget2025} when compared with \ac{MDI} or \ac{SAP}. Nonetheless, as highlighted in a recent review \cite{jacobs2025}, \ac{AID} solutions are not universally appropriate for all people with \ac{T1D}. Their performance is substantially better overnight, when the absence of meals and physical activity simplifies glucose dynamics, than during the more variable conditions of daytime use. 

As current \ac{AID} systems still face important limitations, particularly under the variable conditions of daily life, there is growing interest in therapeutic strategies that can better account for individual differences \cite{hughes2025future}. Therapy personalization offers a promising route to reducing disease burden, attenuating both intra‑ and inter‑subject variability, and identifying behavioral patterns that can be addressed proactively \cite{misra2025ethnic}. Digital precision therapeutics built on \ac{DT} technologies have emerged as a central enabler of this shift. By continuously updating physiologically grounded models with real‑world data, \ac{DT} approaches support prediction, decision support, and adaptive control. Within \ac{T1D}, these technologies are increasingly used to capture subject-specific metabolic dynamics and to drive a broad range of clinical applications~\cite{cappon2024digital}, including real‑time adaptation of \ac{AID} algorithms~\cite{kovatchev2025human}, evaluation of therapeutic interventions~\cite{cappon2023replaybg}, analysis of exercise effects on glucose regulation~\cite{young2024design}, and personalized meal‑bolus recommendations~\cite{builes2025digital}.

At the core of any \ac{DT} implementation lies a computational model capable of reproducing the metabolic behavior of an individual. More broadly, a medical \ac{DT} is a virtual representation of a physical system that is informed by measurements from the real world and, in turn, provides information to support monitoring, simulation, prediction, or therapeutic decision-making~\cite{willcox2023foundational}. Importantly, the computational engine of a \ac{DT} is not restricted to a particular modeling paradigm: it may consist of a purely data-driven model, a physiologically based model, or a hybrid combination of both~\cite{cappon2024digital,cappon2023replaybg,roquemen2026,digitalTwin_Pradigm_dia2026}.

Regardless of the modeling strategy adopted, computational models in medicine can be used to answer different classes of quantitative questions depending on the information available at the time of prediction. Throughout this work, we distinguish between two operational modes of the same individualized model. \textit{Forecasting mode} refers to using the model to forecast glucose over a specified \ac{PH} when current outputs and states are known, but future exogenous inputs are unknown~\cite{liu2019,kushner2020,mosquera2022incorporating,prendin2023,giancotti2024forecasting,basile2025}. In this mode, future glucose trajectories are inferred from the current physiological state and assumptions, estimates, or forecasts of future disturbances. \textit{Intervention-informed mode} refers to using the same model to predict glucose over a specified \ac{PH} when future exogenous inputs are prescribed~\cite{roquemen2026,garcia2018,zhu2023,hoyos2021}. In this mode, the model is used to simulate the physiological response to a defined intervention scenario, such as planned meals, insulin administration, or physical activity. A final distinction is worth emphasizing. Phenomenological models can continuously generate predictions for the variable of interest throughout the entire \ac{PH}. In contrast, most black-box model architectures are designed to provide predictions only at a specific future time point, typically corresponding to the end of the \ac{PH}.

Delivering accurate individualized predictions requires not only an appropriate model structure but also a robust parameter identification strategy. For physiological models, which provide interpretability and mechanistic insight, identification consists of selecting the subset of parameters that most strongly influence the model’s output and estimating their patient-specific values from data. This process is often challenged by limited data, parameter correlations, and structural and practical identifiability issues, which can limit the reliability and uniqueness of the estimated parameter~\cite{balsa2010}. Consequently, global sensitivity analysis has become a standard tool for identifying the subset of parameters that most strongly influence model predictions and should therefore be prioritized for estimation~\cite{sobol-simple-1}. However, extending \ac{SA} to dynamic physiological systems introduces additional challenges, since parameter relevance varies over time and depends on the underlying input conditions~\cite{sobol-time-series-1,sobol-time-series-2,sobol-time-series-3,escorihuela2025parameters}.

In an earlier work~\cite{escorihuela2025parameters}, we introduced a methodology for ranking parameters in compartmental models by adapting Sobol global \ac{SA} to a time‑series setting and explicitly incorporating input variability. However, that initial analysis relied on data from only two individuals (one simulated and one real), limiting the ability to assess inter‑subject variability and the generality of the resulting rankings. While parameter identification is central to individualized therapy, it is equally important to understand whether certain influential parameters remain consistent across individuals. 

In this work, we extend the methodology proposed in~\cite{escorihuela2025parameters} to a large cohort from the \ac{T1DEXI} dataset~\cite{riddell2023examining} to investigate whether a robust, population-level ranking of the \ac{HGM} parameters can be established despite inter-subject variability. Based on this analysis, we identify the subset of parameters to prioritize for identification and evaluate its ability to accurately reproduce subject-specific glucose dynamics. Finally, we assess the predictive performance of the personalized \ac{HGM} under both {interevention-informed} and {forecasting} modes, demonstrating that, when appropriately personalized, physiological models can accurately represent glucose dynamics across both modes, supporting their use as the computational foundation for future digital twin technologies in diabetes.

 The paper is organized as follows. The Methods section outlines the adapted \ac{SA} framework, the compartmental model, the parameter‑selection strategy, the dataset, the identification procedure, and the model evaluation. The Results section reports the outcomes of the \ac{SA} and the parameter identification. The Discussion and Conclusion sections interpret these findings and compare them with those in~\cite{escorihuela2025parameters}.

%% file: Sections/Methods.tex
\section{Methods}
\label{sec:Methods}

\subsection{Hovorka glucoregulatory model for type 1 diabetes}

\begin{figure}
\centering
\includegraphics[trim=0mm 0mm 0mm 0mm, clip=true, width=\linewidth]{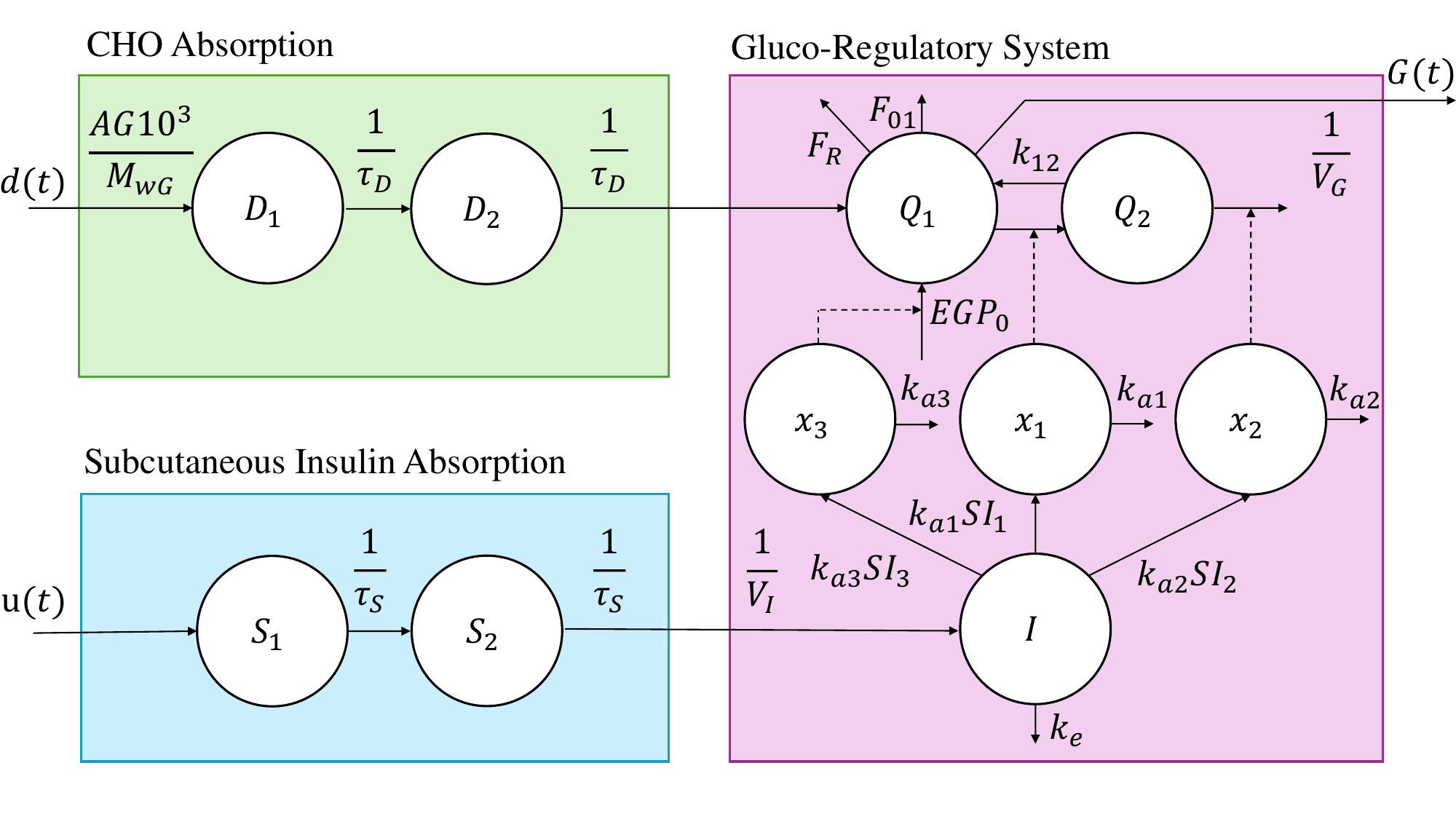}
\caption{Graphical representation of the considered~\ac{HGM}.} 
\label{fig:HovorkaRepresentation}
\end{figure}

Figure \ref{fig:HovorkaRepresentation} shows a schematic representation of the~\ac{HGM} in \ac{T1D} consisting of three sub-models: a first sub-model representing the \ac{CHO} digestion and absorption, the subcutaneous insulin absorption sub-model describing the insulin transport from the subcutaneous tissue to the plasma, and the glucose-insulin system describing glucose regulation in response to physiologic conditions, meal ingestion, and insulin therapy \cite{hovorka-populational-parameters, Wilinska_Hovorka_oralmodel}. This model can be represented as: 
\begin{equation} \label{eq:H1}
\hat{y}(t) = f(t;X(t),U(t),\boldsymbol{\theta})
\end{equation}
with $\hat{y}(t)=G(t)$ the measured glucose concentration in the bloodstream, $X(t)$ the state variables defined in~\eqref{eq:H2}, $U(t)$ the model inputs defined in~\eqref{eq:input-output}, and $\boldsymbol{\theta} \in \mathbb{R}^p$, the vector of $p$ model parameters defined in~\eqref{eq:H4}, and $t$ the time.
\begin{subequations} \label{eq:model_sets}
\begin{align}
X(t) &= [ Q_{1}(t), Q_{2}(t), x_{1}(t), x_{2}(t), x_{3}(t), \nonumber \\[-2pt]
     &\quad S_{1}(t), S_{2}(t), I(t), D_{1}(t), D_{2}(t) ] \label{eq:H2} \\[4pt]
U(t) &= [u(t), d(t)] \label{eq:input-output} \\[4pt]
\boldsymbol{\theta} &= [k_{12}, k_{a1}, k_{a2}, k_{a3}, k_e, AG, \tau_D, \tau_S, \nonumber \\[-2pt]
     &\quad SI_1, SI_2, SI_3, V_I, V_G, EGP_0, F_{01}]\label{eq:H4}
\end{align}
\end{subequations}

\begin{table}[t]
\centering
\scriptsize
\setlength{\tabcolsep}{2.5pt}
\renewcommand{\arraystretch}{1.1}
\caption{Summary of the \ac{HGM} parameters, states, inputs and output.}
\label{tab:HGM_summary}

\begin{tabular*}{\linewidth}{@{\extracolsep{\fill}} l p{3.2cm} p{1cm} l}
\hline
\textbf{Name} & \textbf{Description} & \textbf{Value} & \textbf{Unit} \\
\hline
\multicolumn{4}{l}{\textbf{Parameters}} \\
\hline
$k_{12}$ & Transfer rate & 0.066 & min$^{-1}$ \\
$EGP_0$ & EGP at zero insulin & 0.0161 & mmol\,kg$^{-1}$\,min$^{-1}$ \\
$k_{a1}$ & Deactivation rate (transport) & 0.006 & min$^{-1}$ \\
$SI_1=\tfrac{k_{b1}}{k_{a1}}$ & Insulin sensitivity (transport) & $51.2{\times}10^{-4}$ & min$^{-1}$ L mU$^{-1}$ \\
$k_{a2}$ & Deactivation rate (disposal) & 0.06 & min$^{-1}$ \\
$SI_2=\tfrac{k_{b2}}{k_{a2}}$ & Insulin sensitivity (disposal) & $8.2{\times}10^{-4}$ & min$^{-1}$ L mU$^{-1}$ \\
$k_{a3}$ & Deactivation rate (EGP) & 0.03 & min$^{-1}$ \\
$SI_3=\tfrac{k_{b3}}{k_{a3}}$ & Insulin sensitivity (EGP) & $520{\times}10^{-4}$ & min$^{-1}$ L mU$^{-1}$ \\
$\tau_S$ & Insulin absorption constant & 55 & min \\
$k_e$ & Insulin elimination rate & 0.138 & min$^{-1}$ \\
$AG$ & CHO availability & 0.8 & — \\
$M_{wg}$ & Glucose molecular weight & 180.16 & g\,mol$^{-1}$ \\
$\tau_D$ & CHO absorption time constant & 40 & min \\
$V_G$ & Glucose distribution volume & 0.16 & L\,kg$^{-1}$ \\
$V_I$ & Insulin distribution volume & 0.12 & L\,kg$^{-1}$ \\
$F_{01}$ & Non-insulin glucose flux & 0.0097 & mmol\,kg$^{-1}$\,min$^{-1}$ \\
\hline
\multicolumn{4}{l}{\textbf{States}} \\
\hline
$Q_{1}(t)$ & \multicolumn{2}{p{4.7cm}}{Glucose mass, accessible compartment} & mmol\,kg$^{-1}$ \\
$Q_{2}(t)$ & \multicolumn{2}{p{4.7cm}}{Glucose mass, non-accessible compartment} & mmol\,kg$^{-1}$ \\
$x_{1}(t)$ & \multicolumn{2}{p{4.7cm}}{Insulin effect on glucose transport} & min$^{-1}$ \\
$x_{2}(t)$ & \multicolumn{2}{p{4.7cm}}{Insulin effect on glucose disposal} & min$^{-1}$ \\
$x_{3}(t)$ & \multicolumn{2}{p{4.7cm}}{Insulin effect on \ac{EGP}} & min$^{-1}$ \\
$S_{1}(t)$ & \multicolumn{2}{p{4.7cm}}{Subcutaneous insulin compartment 1} & mU\,kg$^{-1}$ \\
$S_{2}(t)$ & \multicolumn{2}{p{4.7cm}}{Subcutaneous insulin compartment 2} & mU\,kg$^{-1}$ \\
$I(t)$ & \multicolumn{2}{p{4.7cm}}{Plasma insulin concentration} & mU\,L$^{-1}$ \\
$D_{1}(t)$ & \multicolumn{2}{p{4.7cm}}{CHO absorption compartment 1} & mmol\,kg$^{-1}$ \\
$D_{2}(t)$ & \multicolumn{2}{p{4.7cm}}{CHO absorption compartment 2} & mmol\,kg$^{-1}$ \\

\hline
\multicolumn{4}{l}{\textbf{Inputs}} \\
\hline
$u(t)$ & \multicolumn{2}{p{4.7cm}}{Insulin infusion rate} & mU\,min$^{-1}$ \\
$d(t)$ & \multicolumn{2}{p{4.7cm}}{Carbohydrate intake rate} & g\,min$^{-1}$ \\

\hline
\multicolumn{4}{l}{\textbf{Output}} \\
\hline
$G(t)$ & \multicolumn{2}{p{4.7cm}}{Plasma glucose concentration} & mmol\,L$^{-1}$ \\
\hline
\end{tabular*}

\end{table}

The \ac{HGM} equations considered in this study are provided in  Appendix~\ref{sec:appendix Hovorka} and the details of the model states~\eqref{eq:H2}, inputs~\eqref{eq:input-output}, parameters~\eqref{eq:H4}, and output $G(t)$ are described in Table~\ref{tab:HGM_summary}.

\subsection{Sobol sensitivity analysis}\label{sec:sobol sens analysis}

The Sobol method is a global \ac{SA} technique widely used in computational modeling to estimate the influence of model parameters on the output variance \cite{sensitivity-analysis-the-primer-book}. The method, based on variance decomposition, quantifies the contribution of each parameter $\theta_i \in \boldsymbol{\theta}$ to the variance of the model output $\hat{y}(t)$ by decomposing the total output variance into components associated with individual parameters and their interactions. 

Let us assume $\boldsymbol{\theta} \in \Omega_p \subseteq \mathbb{R}^p$, a random vector of $p$ statistically independent model parameters with a uniform distribution spanning $\Omega_p$. The total output variance can be decomposed as follows:
\begin{equation} \label{eq:varianceDecompositionMain}
\begin{split}
\text{Var}(f,t) &= \sum_{i=1}^{p} V_{\theta_i}(f,t) + \sum_{i<j}^{p} V_{\theta_i,\theta_j}(f,t) + \dots \\
&+ V_{\theta_1,\theta_2,\dots,\theta_{p}}(f,t) 
\end{split}
\end{equation}
where $\text{Var}(\cdot)$ is the variance operator and \( V_{\theta_i} \) and \( V_{\theta_i,\theta_j} \) are defined as:   
\begin{subequations} \label{eq:varianceDecomposition}
\begin{align}
V_{\theta_i}(f,t) = &\text{Var}_{\theta_i}\left( \mathbb{E}_{\theta_{\sim i}}[f,t\vert\theta_i] \right) \label{eq:varianceDecomposition1} \\
V_{\theta_i,\theta_j}(f,t) = &\text{Var}_{\theta_i,\theta_j}\left( \mathbb{E}_{\theta_{\sim i},\theta_{\sim j}}[f,t\vert\theta_i, \theta_j] \right) \label{eq:varianceDecomposition2} - \dots \\
&- V_{\theta_i}(f,t) - V_{\theta_j}(f,t) \nonumber
\end{align}
\end{subequations}
with $\text{Var}_{\theta_i}$ representing the variance caused by the parameter $\theta_i$. $\mathbb{E}_{\theta_{\sim i}}[f,t\vert\theta_i]$ and $\mathbb{E}_{\theta_{\sim i},\theta_{\sim j}}[f,t\vert\theta_i, \theta_j]$ denote the conditional expectations of the output with respect to all parameters except $\theta_i$ and except $\theta_i$ and $\theta_j$, respectively. 

Sobol’ indices are typically reported as scalar quantities, providing single, time-independent assessments \cite{sensitivity-analysis-the-primer-book,saltelli2010variance, azzini2021sobol, kucherenko2017different, owen2013better, sobol-simple-1}. A way to dynamically distinguish them is by using the pointwise-in-time first- ($S_{\theta_i}(f,t)$) and total- ($S_{\theta_i}^T(f,t)$) order Sobol’ indices, enabling dynamic sensitivity analyses of model parameters over time \cite{randall2021}:
\begin{subequations}
\label{eq:Si}
\begin{align}
S_{\theta_i}(f,t)
&=
\frac{V_{\theta_i}(f,t)}
{\mathrm{Var}(f,t)}
\label{eq:sobolfirst}
\\
S_{\theta_i}^{T}(f,t)
&=
1-
\frac{V_{\theta_{\sim i}}(f,t)}
{\mathrm{Var}(f,t)}
\label{eq:soboltotal}
\end{align}
where $V_{\theta_{\sim i}}(f,t)$ is defined as:
\begin{equation}
\label{eq:varianceTotal}
V_{\theta_{\sim i}}(f,t)
=
\mathrm{Var}_{\theta_{\sim i}}
\left(
\mathbb{E}_{\theta_i}
\left[
f,t\,\middle|\,\theta_{\sim i}
\right]
\right).
\end{equation}
\end{subequations}

Here, $\mathrm{Var}_{\theta_{\sim i}}(\cdot)$ denotes the variance computed over the complementary parameter set $\theta_{\sim i}$, i.e., all parameters except $\theta_i$, while $\mathbb{E}_{\theta_i}[f,t\,|\,\theta_{\sim i}]$ denotes the conditional expectation of the model output with respect to $\theta_i$, conditioned on the values of all complementary parameters.

Conceptually, $S_{\theta_i}(f,t)$ represents the fraction of the output variance explained by the direct effect of a single parameter $\theta_i$ as a function of time. In contrast, $S_{\theta_i}^T(f,t)$ captures both the individual effects of the parameter and all its interactions with other parameters in the light of the model structure, also as a function of time.

In this study, we adapted the method proposed in \cite{sobol-simple-1} to obtain dynamic estimates of the Sobol' sensitivity indices for a time-varying system, motivated by the following considerations. Let us consider a model $f$ with time-varying output $\hat{y}(t)$ as in \eqref{eq:H1} on the interval $[0,N_D]$, with $N_D>0$, i.e., $\hat{y}(t)=f(t;X(t),U(t),\theta)$ for $t\in [0,N_D]$. 

Computing Sobol’ indices using standard methods, e.g., based on \eqref{eq:sobolfirst} and \eqref{eq:soboltotal}, poses two main challenges. First, classical \ac{SA} analysis yields $p$ time series in $t$, with $t=1, \dots, N_{D}$ consisting of pointwise-in-time Sobol’ indices, one for each sampling time and each parameter. Since the output variance often varies over time, direct comparison of these $N_{D}$ indices may fail to capture the dynamic effects induced by different inputs, the system's memory, and the resulting between-time correlations. Consequently, defining a single summary metric that meaningfully represents the overall contribution of a parameter over $[0, N_D]$ is non-trivial.
Second, Sobol’ indices are highly sensitive to operating conditions, and the relative importance of parameters may vary substantially across different input trajectories. In receding-horizon counterfactual prediction, such as that performed by a Model Predictive Controller, the \ac{PH} is often much shorter than the length of the available input–output dataset (i.e., $N_{PH} \ll N_{D}$). Therefore, performing \ac{SA} without considering the entire input-output sequence may overlook instances in which certain parameters are influential only under certain conditions.

Solutions to expand the \ac{SA} for time-dependent systems have been provided elsewhere \cite{sobol-time-series-2, randall2021}. In this work, we adopt an approach similar to that introduced by Randall et al. \cite{randall2021} for computing limited-memory Sobol’ indices. First, we normalize the Sobol' indices over a given prediction horizon to ensure a balanced and interpretable sensitivity contrast. This approach yields the weighted first- and total-order Sobol index $S_{\theta_i}^w$ and $S_{\theta_i}^{T,w}$ over $k\in [t,t+N_{PH}]$, respectively, as:
\begin{subequations} \label{eq:SiNorms}
\begin{align}
S_{\theta_i}^w(f,t) &= \frac{1}{N_{PH}+1} \sum_{k=t}^{t+N_{PH}} w_{k} S_{\theta_i}(f,k) \label{eq:SiNorm}\\
S_{\theta_i}^{T,w}(f,t) &= \frac{1}{N_{PH}+1} \sum_{k=t}^{t+N_{PH}} w_{k} S_{\theta_i}^T(f,k) \label{eq:SiTotNorm}
\end{align}
with $w_{k}$ defined as:
\begin{equation} \label{eq:weight}
w_{k} = \frac{\text{Var}(f,k)}{\max\limits_{k\in[t,~t+N_{PH}]} \text{Var}(f,k)}, \quad w_{k}\in[0,~1]
\end{equation}
\end{subequations}

Bearing the above considerations in mind, we propose Algorithm~\ref{algo:sens indices and ranking} to quantify the influence of model parameters in the model structure \eqref{eq:H1} while accounting for the full input-output sequence $\{U(t), \hat{y}(t)\}_{t=1}^{N_D}$. This approach is designed to account for multiple dynamic conditions, rather than a single fixed scenario.

Algorithm~\ref{algo:sens indices and ranking} provides $S_{\theta_i}^w(f,t)$ and $S_{\theta_i}^{T,w}(f,t)$ up to $t \in [1,N_D-N_{PH}]$. The algorithm outputs are two vectors, $\boldsymbol{v^{S_{\theta_i}}} \in \mathbb{R}^{N_D - N_{PH}}$ and $\boldsymbol{v^{S_{\theta_i}^T}}  \in \mathbb{R}^{N_D - N_{PH}}$, representing the temporal evolution of the first- and total-order sensitivity indices for each parameter across the entire input–output sequence.

\begin{algorithm}[htpb!]
\caption{Estimation of parametric sensitivity indices in dynamical systems.}
\label{algo:sens indices and ranking}

\textbf{Input}: Model \eqref{eq:H1}, input sequence $\{U(t)\}_{t=1}^{N_D}$, and sampled realizations of $\theta$. %
\vspace*{.1cm}\hrule\vspace*{.1cm}

    \begin{algorithmic}[1]

    \For{$t = 1 ~\text{to}~ N_D -N_{PH}$}
      
        \State  Compute model predictions for $N_{PH}$ samples using the given physiological model, sampled realizations of $\theta$, and inputs $\{U(k)\}_{k={t}}^{t+{N_{PH}}}$ \label{step:prediction}
    
        \For{$k = t \ \text{to} \ t+N_{PH}$}
            \State Compute $S_{\theta_i}(f,k)$ using~\eqref{eq:sobolfirst}
            \label{step:S_t,theta}
            \State Compute $S_{\theta_i}^T(f,k)$ using~\eqref{eq:soboltotal}
            \label{step:S^T,theta}
        \EndFor
    
        \State Compute $S_{\theta_i}^w(f,t)$ using~\eqref{eq:SiNorm}
        \State Compute $S_{\theta_i}^{T,w}(f,t)$ using~\eqref{eq:SiTotNorm}
        \State $\boldsymbol{v^{S_{\theta_i}}}(t) \gets S_{\theta_i}^w(f,t)$ and  $\boldsymbol{v^{S_{\theta_i}^T}}(t) \gets S_{\theta_i}^{T,w}(f,t)$
    \EndFor
    \end{algorithmic}

\vspace*{.1cm}\hrule\vspace*{.1cm}
\textbf{Output}: $\boldsymbol{v^{S_{\theta_i}}} \in \mathbb{R}^{(N_D - N_{PH})}$ and $\boldsymbol{v^{S_{\theta_i}^T}} \in \mathbb{R}^{(N_D - N_{PH})}$ for each $\theta_i \in \boldsymbol{\theta}$. 
\end{algorithm}

Finally, the average first- and total-order Sobol' indices $\bar{S}_{\theta_i}$ and $\bar{S}_{\theta_i}^T$ over $[1,N_D]$ are computed as:

\begin{subequations}\label{eq:rankingAvg}
\begin{align}
\bar{S}_{\theta_i}
&=
\frac{1}{N_D-N_{PH}}
\sum_{t=1}^{N_D-N_{PH}}
\boldsymbol{v^{S_{\theta_i}}}(t)
\label{eq:rankingFirst}
\\
\bar{S}_{\theta_i}^{T}
&=
\frac{1}{N_D-N_{PH}}
\sum_{t=1}^{N_D-N_{PH}}
\boldsymbol{v^{S_{T_{\theta_i}}}}(t)
\label{eq:rankingTotal}
\end{align}
\end{subequations}

where
\[
\bar{S}_{\theta_i},\;
\bar{S}_{\theta_i}^{T}
\in [0,1].
\]

These metrics, bounded between 0 and 1, quantify the average influence of each parameter on the model output across the entire input-output sequence $\{U(t),\hat{y}(t)\}_{t=1}^{N_D}$, where 0 indicates no influence, and 1 represents maximum influence. 

It is important to note that Algorithm~\ref{algo:sens indices and ranking} can be applied in the intervention-informed and forecasting modes, depending on the design decision. The distinction arises in Step~\ref{step:prediction}, where the input information available to the model during prediction is specified. %

\subsection{The T1DEXI adult cohort dataset}
\label{S2.5}
Numerical analyses were conducted using data from adult \ac{T1D} participants of the \ac{T1DEXI} dataset under closed-loop therapy~\cite{riddell2023examining}. The dataset included \ac{CGM} measurements, bolus and basal insulin infusion records, and carbohydrate intake, all sampled at 5-minute intervals. Of the 216 participants using closed-loop therapy, we included only those with at least 10 consecutive days of data, where each day contained less than 20\% missing values in either \ac{CGM} readings or basal insulin infusion records. Missing segments shorter than 20 minutes were linearly interpolated. This selection resulted in a cohort of 192 participants.  
Finally, for each participant, the resulting 10-day segments were partitioned into two datasets: 7 days for the parameter identification and 3 days for the validation. Within the model identification dataset, a 1-day window was used to perform the \ac{SA} and obtain a participant ranking of the model parameters. Demographic characteristics and glycemic metrics for the included participants are summarized in Table~\ref{tab:participant_summary}.

\begin{table}[htpb!]
\renewcommand{\arraystretch}{1.3}
\caption[Participant characteristics and glycemic outcomes (192 participants, 10 days)]{Participant characteristics and glycemic outcomes (192 participants, 10 days). Metrics are reported as mean±SD if normally distributed. Non-normal continuous metrics are reported as median [IQR].}
\label{tab:participant_summary}

\begin{tabular}{p{5.5cm} p{2cm}}
\hline
\textbf{Variable} & \textbf{Value} \\
\hline
Female (\#N/\%) & 144/75.4 \\
Age (years) & $40 \pm 15$ \\
Diabetes duration (years) & $21.6 \pm 14.8$ \\
BMI (kg/m$^2$) & $25.4 \pm 3.9$ \\
Total daily basal insulin (U/day) & $20.1 \pm 10.6$ \\
Total daily bolus insulin (U/day) & $20.8 \pm 11.6$ \\
Total daily insulin (U/day) & $40.9 \pm 19.4$ \\
Total daily insulin per kg(U/kg/day) & $0.6 \pm 0.2$ \\
CHO intake (g/day) & $128~[90,170]$ \\
Coefficient of variation (\%/day) & $27.9 \pm 7.4$ \\
Time in [70-180] mg/dL (\%)& $78.4 \pm 15.6$ \\
Time below 70 mg/dL (\%) & $2.1 \pm 3.5$ \\
Time above 180 mg/dL (\%) & $17.6 \pm 15.2$ \\
Average glucose (mg/dL) & $141.4 \pm 22.1$ \\
\hline
\end{tabular}
\end{table}

\subsection{Global parameter ranking}
\label{subsec:parameterRanking}

This section describes the procedure to obtain a global ranking for the parameters of the~\ac{HGM} after applying Algorithm~1 using insulin, meal, and \ac{CGM} data collected from participants in the \ac{T1DEXI} dataset.
As described in Section
~\ref{sec:sobol sens analysis}, Algorithm~1 uses the~\ac{HGM} model \eqref{eq:H1} along with an input-output sequence $\{U(t), \hat{y}(t)\}_{t=1}^{N_D}$ containing $N_D$ samples of insulin, carbohydrate intake, and \ac{CGM} data to produce $\boldsymbol{v^{S_{\theta_i}}}$ and $\boldsymbol{v^{S_{T_{\theta_i}}}}$, the aggregated first- and total-order sensitivity indices for every $\theta_i$ in the entire evaluated input-output sequence. In this study, the \ac{SA} was performed under the intervention-informed mode, to ensure that the parameters capture the complete glucose-insulin dynamics associated with the evaluated disturbances.

For each participant, Algorithm~1 was applied to a representative day containing at least (i) one meal without a prandial bolus, (ii) one isolated insulin bolus, and (iii) one postprandial episode involving both a meal and a prandial bolus. The criteria defining these dynamic events are summarized in Table~\ref{tab:characteristics_sensitivity_analysis_day}. %

To characterize parameter importance under different physiological excitation patterns, four parameter rankings were computed: an \say{all day} ranking using the complete input-output sequence, together with three event-specific rankings corresponding to \say{meal only}, \say{bolus injection}, and \say{postprandial response} conditions.

For the \say{all day} analysis, the aggregated first-order sensitivity index $\bar{S}_{\theta_i}$ was computed over the entire input-output sequence using~\eqref{eq:rankingFirst}, and parameters were ranked according to the average $\bar{S}_{\theta_i}$ across participants. Parameter rankings were based on first-order Sobol indices, as we evaluate the isolated contribution of each parameter to the model output. Higher-order indices capturing parameter interactions were omitted, since investigating interaction effects and identifiability issues lies outside the scope of this study.

For the three event-specific analyses, $\bar{S}_{\theta_i}$ was computed within a 20-minute window centered on each event. Restricting the analysis to these intervals enables the rankings to capture parameter relevance during well-defined physiological responses rather than averaging their influence over the entire day. Parameters were subsequently ranked according to the average value of $\bar{S}_{\theta_i}$ across participants.

\begin{table}[H]
\centering
\caption{Input characteristics used to select the day for \ac{SA}. Each condition had to be satisfied within a 20-minute window. $\checkmark$ and \ding{55} indicate whether the corresponding input was present or absent, respectively.}
\label{tab:characteristics_sensitivity_analysis_day}

\begin{tabularx}{\columnwidth}{
    >{\raggedright\arraybackslash}X
    >{\centering\arraybackslash}X
    >{\centering\arraybackslash}X
    >{\centering\arraybackslash}X
}
\hline
\textbf{Scenario} & \textbf{CHO intake} & \textbf{Bolus dose} & \textbf{Basal rate} \\
\hline
Meal only
& $\checkmark$
& \ding{55}
& $\checkmark$/\ding{55} \\

Bolus inj.
& \ding{55}
& $\checkmark$
& $\checkmark$/\ding{55} \\

Postprandial
& $\checkmark$
& $\checkmark$
& $\checkmark$/\ding{55} \\
\hline
\end{tabularx}

\end{table}

Finally, the four rankings were combined into a single global ranking using a weighted aggregation strategy. Under each analysis, every parameter was assigned a rank $r_{i,d}\in{1,\ldots,p}$, with lower ranks indicating greater importance. The global score for parameter $\theta_i$ was then computed as:
\begin{equation}
R_i=\sum_{d=1}^{D} h_d \cdot r_{i,d}, \qquad i=1,\ldots,p
\label{eq}
\end{equation}

where $D=4$ denotes the number of evaluated dynamic conditions, $r_{i,d}$ is the rank of parameter $\theta_i$ under condition $d$, and $h_d\geq0$ is the weight assigned to that condition.

In this study, equal weights ($h_d=1$) were assigned to all four rankings. Because each condition activates distinct physiological mechanisms in the \ac{HGM}, they provide complementary information about parameter relevance. Equal weighting, therefore, yields a global ranking that reflects parameter importance across the full range of clinically relevant glucose-insulin dynamics, rather than emphasizing any single condition. More generally, the weights $h_d$ can be adjusted to prioritize specific dynamic conditions depending on the intended application. The final global ranking is obtained by sorting the parameters in descending order of $R_i$, with higher scores indicating greater overall importance.

\begin{figure*}[t]  %
     \centering
     \includegraphics[trim=25mm 0mm 20mm 0mm, clip=true, width=\textwidth]{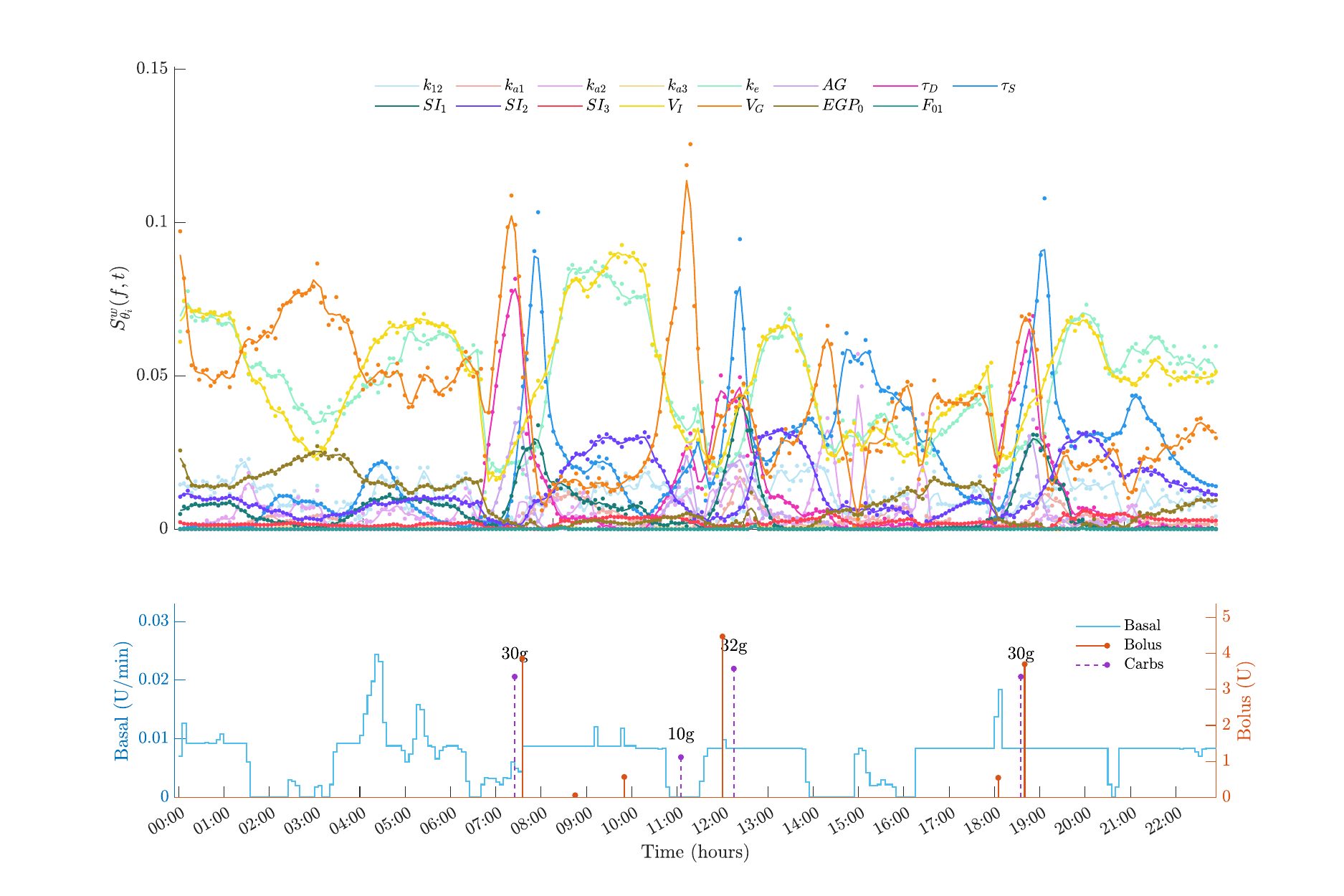}   
     \caption{$S_{\theta_i}^w(f,t)$ indices over one day of data for a selected participant from \ac{T1DEXI} dataset. The top panel shows $S_{\theta_i}^w(f,t)$ indices computed with $f$ from Eq.~\eqref{eq:H1} at every $t=[1,N_D-N_{PH}]$ sampling time for 1-hour predictions. The bottom panel represents insulin and carbohydrate intake input data.} 
     \label{fig:first_indices_all_day}
\end{figure*}

 \begin{figure*}[!t]
 \centering
 \subfigure[All day.\label{fig:ranking_all_day}]{ \includegraphics[width=0.45\textwidth]{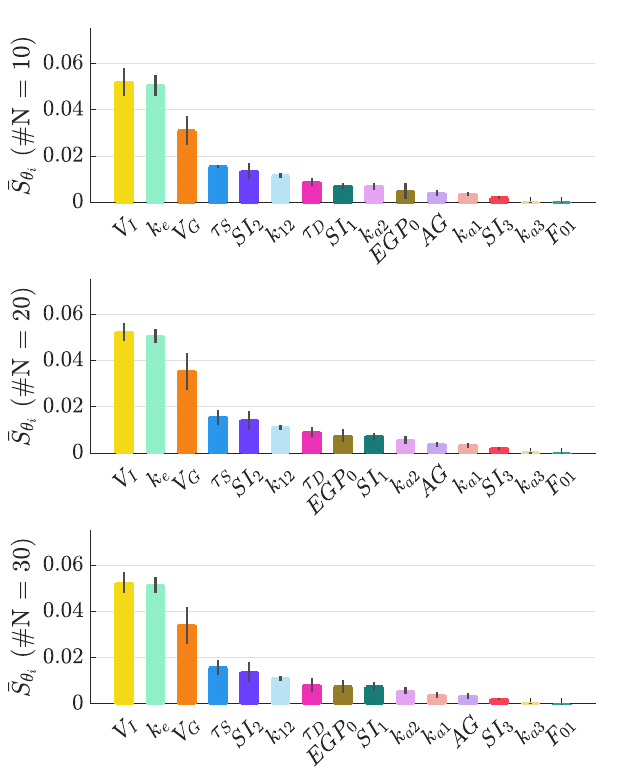} }
 \hfill 
 \subfigure[Meal only.\label{fig:ranking_meal}]{ \includegraphics[width=0.45\textwidth]{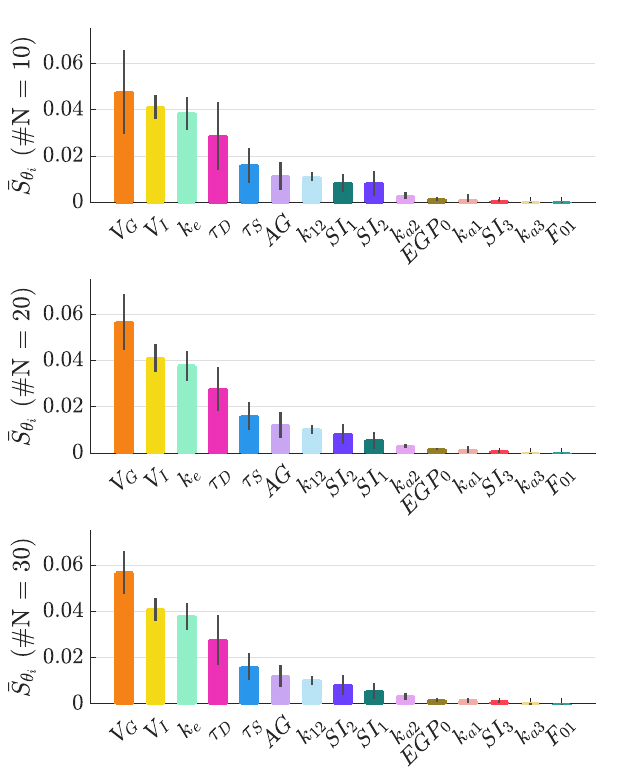} } \medskip 
 \subfigure[Bolus injection.\label{fig:ranking_bolus}]{ \includegraphics[width=0.45\textwidth]{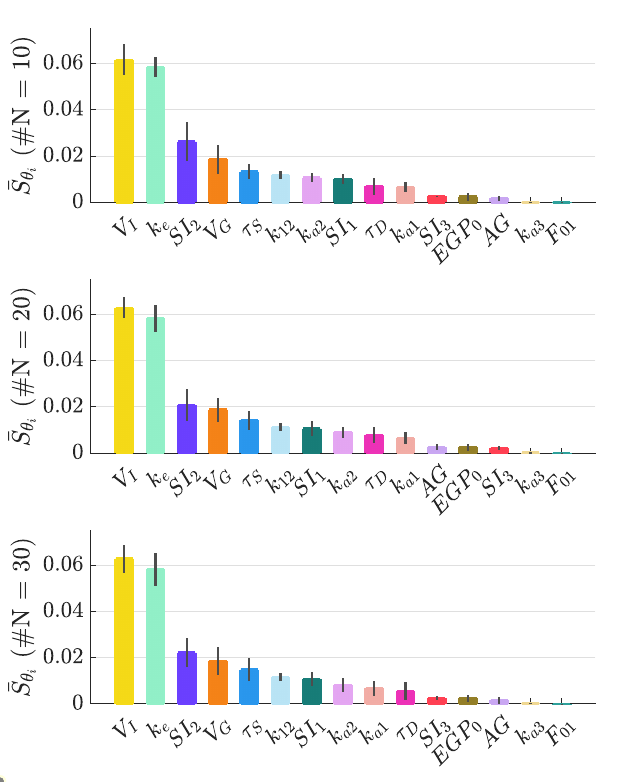} }
 \hfill
 \subfigure[Postprandial response.\label{fig:ranking_pp}]{ \includegraphics[width=0.45\textwidth]{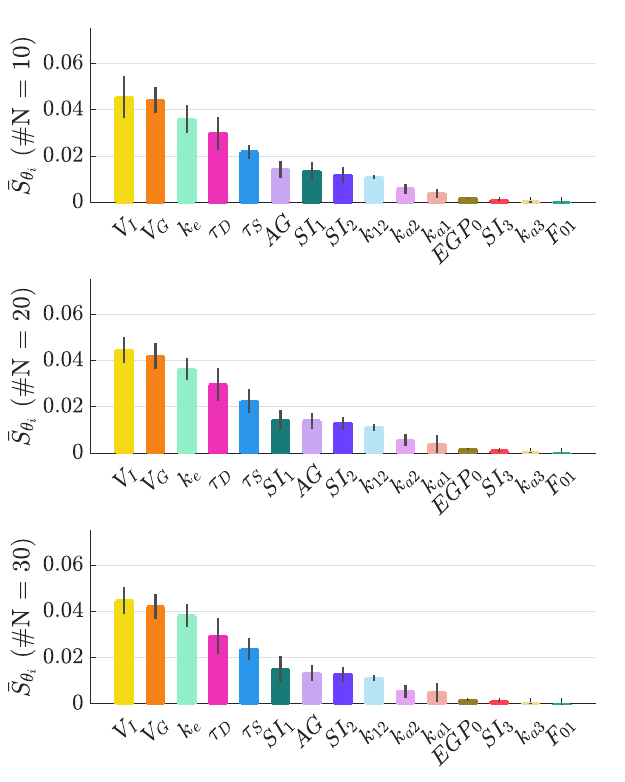} } 
 \caption{ Ranking of parameters obtained using Algorithm~1 for $\#$N=10, $\#$N=20, and $\#$N=30 participants. Subfigure (a) shows the ranking across the full dataset, while subfigures (b), (c), and (d) correspond to "meal only", "bolus injection", and "postprandial response", respectively. The criteria defining these events are provided in Table~\ref{tab:characteristics_sensitivity_analysis_day}. } \label{fig:ranking_parameters_30_participants} 
 \end{figure*}

\begin{figure}[!t]
     \centering
     \includegraphics[width=\linewidth]{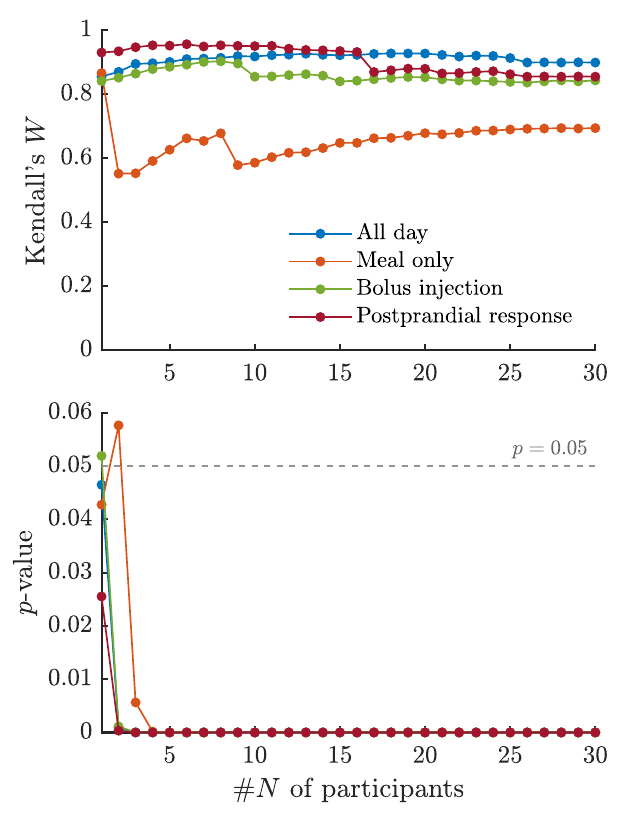}
     \caption{Kendall's W test and respective p-value to assess statistical significance in ranking concordance while adding one participant at each step.}
     \label{fig:statisticalTest}
\end{figure}

\subsection{Parameter identification}

Parameter identification was performed independently for each participant. The subset of parameters to be identified was selected according to the proposed global parameter ranking.

A nonlinear least-squares optimization algorithm was used to identify the model parameters by minimizing the cost function $\mathcal{J}$, defined as the average Root Mean Square Error ($\textit{RMSE}_{av}$) over all prediction windows:

\begin{subequations}\label{eq:RMSE}
\begin{flalign}
\textit{RMSE}(t) &=\sqrt{
\frac{1}{N_{PH}}
\sum_{k=0}^{N_{PH}-1}
\left[y(t+k)-\hat{y}(t+k)\right]^2}
&&\label{eq:RMSE_window}
\\[0.8em]
\textit{RMSE}_{av} &= 
\frac{1}{N_D-N_{PH}}
\sum_{t=1}^{N_D-N_{PH}}
\textit{RMSE}(t)
&&\label{eq:RMSE_total}
\end{flalign}
\end{subequations}

where $y(t+k)$ and $\hat{y}(t+k)$ denote the measured and predicted glucose concentrations, respectively.

For each participant, the parameter identification was performed using the {intervention-informed} mode, where the complete future input sequence was assumed to be available. The input-output dataset was defined as $\{U(t),\hat{y}(t)\}_{t=1}^{N_D}$. At every sampling instant $t=1,\ldots,N_D-N_{PH}$, a 60-min prediction trajectory,
$\{\hat{y}(t)\}_{t=n_d}^{n_d+N_{PH}}$, was generated using~\eqref{eq:H1} along with the corresponding future input sequence
$\{U(t)\}_{t=n_d}^{n_d+N_{PH}}$. In this work, $N_D=2016$ samples (7 days) and $N_{PH}=12$ samples (60 min).

\subsection{Model performance analysis}
Model performance was evaluated using the $\textit{RMSE}_{av}$ defined in~\eqref{eq:RMSE}. In addition, the endpoint prediction error at PHs of 15, 30, and 60 min was quantified using $\mathrm{RMSE}_{N_{PH}}$, where $N_{PH}$ denotes the number of samples corresponding to the \ac{PH}. $\mathrm{RMSE}_{N_{PH}}$ is defined as:

\begin{subequations} \label{eq:rmse_fix_horizon}
\begin{align}
\textit{RMSE}_{N_{PH}}
&=
\sqrt{
\frac{1}{N_D-N_{PH}}
\sum_{t=1}^{N_D-N_{PH}}
\varepsilon_t^2
}
\end{align}
where,
\begin{equation}
\varepsilon_t
=
y(t+N_{PH})
-\hat{y}(t+N_{PH})
\end{equation}
\end{subequations}

Here, $y(t+N_{PH})$ and $\hat{y}(t+N_{PH})$ denote the measured and predicted glucose concentrations at the end of the \ac{PH} for a prediction initiated at time $t$, respectively. The latest is the most commonly reported metric in glucose prediction literature for evaluating models in {forecasting} modes~\cite{liu2019,kushner2020,zhu2022}.

Performance was assessed under both the {intervention-informed} and {forecasting} modes. In the {intervention-informed} mode, the true future insulin and meal inputs were provided to the model. In contrast, the {forecasting} mode assumed no knowledge of future disturbances: future bolus injections and carbohydrate intakes were set to zero, while basal insulin delivery was assumed to remain constant at its current value throughout the \ac{PH}.

%% file: Sections/Results.tex
\section{Results} \label{sec:Results}
\subsection{Global parameter ranking}

We applied Algorithm 1 to determine the most influential parameters of the \ac{HGM} given real-life excitation conditions. Since estimating Sobol' sensitivity indices is computationally intensive, we started with a subset of 10 participants and analyzed how the parameter rankings changed as we added 10 participants at a time until reaching 30. To quantify the agreement in parameter rankings across participants, the Kendall's coefficient of concordance (Kendall's W) was computed incrementally by adding one participant at a time up to the full cohort of 30 participants for each evaluated event-specific condition.

Algorithm 1 used the input sequence $\{u(t), d(t)\}_{t=1}^{N_D}$ over one day of data, resulting in $N_D = 288$ samples. Glucose predictions were computed using model~\eqref{eq:H1} over a 1-hour \ac{PH}, corresponding to $N_{PH} = 12$ prediction samples. State estimates were obtained via an \ac{EKF}.
Parameters of the \ac{EKF} are detailed in  Appendix~\ref{sec:appendix EKF}.

$S_{\theta_i}(f,t)$ were computed following~\cite{sobol-simple-1}, using 3000 glucose predictions per parameter and a Latin Hypercube Sampling (LHS) scheme, with parameter ranges defined as $[0.1~ 2] \times \theta_{i,n}$, with $\theta_{i,n}$ the nominal value for $\theta_{i}$. These limits were selected to match those used in the subsequent identification strategy.  

Figure~\ref{fig:first_indices_all_day} shows the Sobol' \ac{SA} for one representative participant from the \ac{T1DEXI} dataset. The top panel presents the weighted first-order Sobol' index sequences for every parameter in the considered dataset. 
The bottom panel shows basal and bolus insulin injections, as well as carbohydrate intake. Single bolus injections occur around 09:00h, 10:00h, and 06:00h, isolated meals right after 11:00h, and postprandial responses around 07:30h, 12:00h, and 18:30h.

Figure~\ref{fig:ranking_parameters_30_participants} shows the ranking of parameters for the different considered conditions \say{all day}, \say{meal only}, \say{bolus injection}, and  \say{postprandial response} across $\#$N participants, with $\#$N$=\{10, 20, 30\}$, where parameters were ranked in descending order according to the median of $\bar{S}_{\theta_i}$. Specifically, Figure~\ref{fig:ranking_all_day} shows the ranking of the sensitivity indexes over the \say{all day}, and Figure~\ref{fig:ranking_meal} to~\ref{fig:ranking_pp} show the ranking for the \say{meal only}, \say{bolus injection}, and \say{postprandial response}, respectively. 

The parameter ranking remains remarkably consistent as the number of participants increases, indicating that the relative importance of the parameters is stable across individuals. This observation is further supported by the Kendall's W analysis in Figure~\ref{fig:statisticalTest}, where both the concordance coefficient and the associated p-values demonstrate a high level of agreement in the rankings from the earliest participants onward. The volume of insulin distribution ($V_I$) and insulin elimination rate from plasma ($k_e$) remain the two key parameters with the highest $\bar{S}_{\theta_i}$ values. However, their importance decreases during postprandial responses, in which the glucose distribution volume ($V_G$), the carbohydrate absorption constant ($\tau_D$), and the insulin absorption constant ($\tau_S$) become more relevant.

We derive a single global ranking that accounts for inter-subject variability across all dynamic conditions, applying the majority-voting strategy described in Section~\ref{subsec:parameterRanking} to the 30 participants, yielding:

\begin{equation*}
\begin{split}
\boldsymbol{\theta}^{H}_{rank} = [&V_I, k_e, V_G, \tau_S, SI_2, \tau_D, k_{12}, SI_1, k_{a2}, \\ &AG, EGP_0, k_{a1}, SI_3, k_{a3}, F_{01}]
\end{split}
\end{equation*}

as the resulting global parameter ranking.

\subsection{Parameter identification and model performance analysis} 

Following the global parameter ranking, we quantified the improvement in prediction performance obtained by progressively expanding the set of identified parameters. Parameters were introduced sequentially according to their global ranking, and the identification procedure was terminated once the $\textit{RMSE}_{av}$ in \eqref{eq:RMSE_total} reached a plateau, indicating that the inclusion of additional parameters no longer produced meaningful improvements in predictive performance.

\begin{figure*}[!t] %
    \centering
    \includegraphics[trim=2mm 0mm 5mm 0mm, clip=true, width=\textwidth]{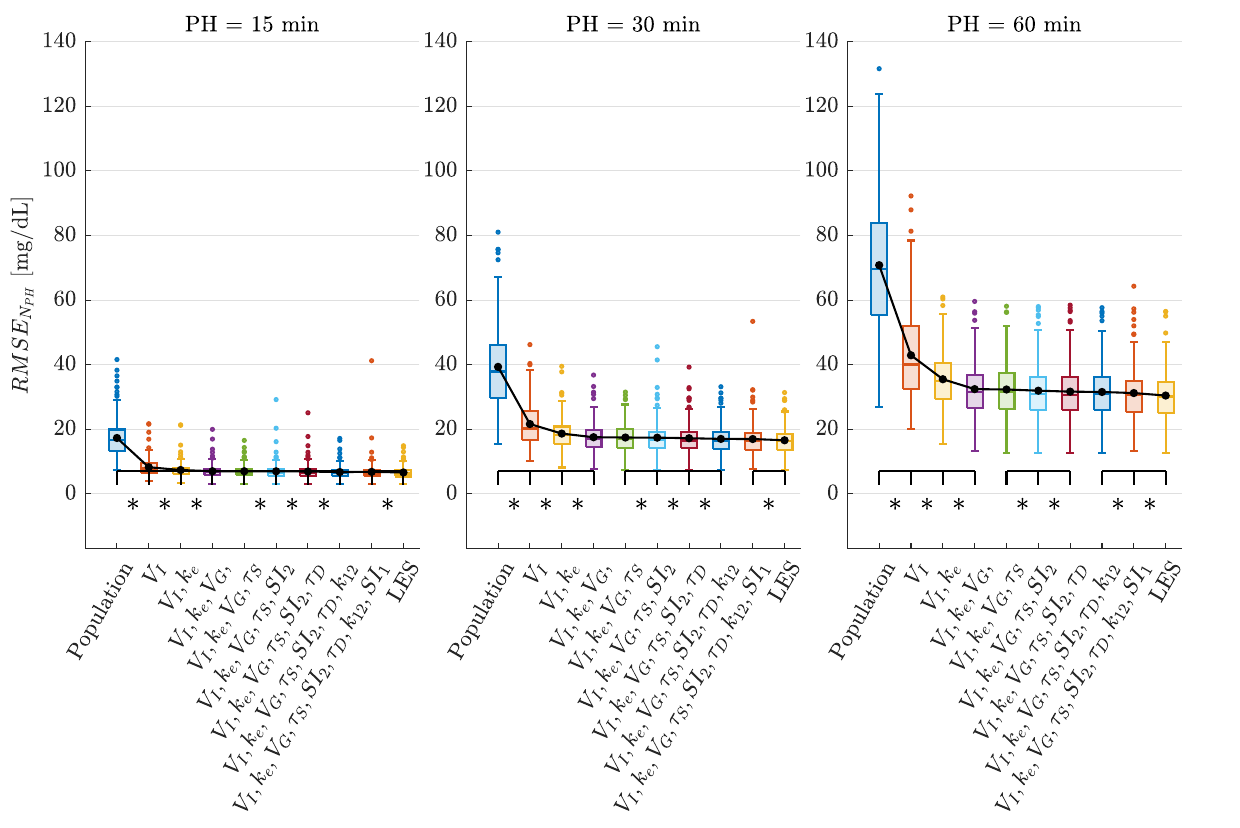}
    \caption{Distribution of validation $RMSE_{N_{PH}}$ values for all 192 participants in the "all-day" case at 15-, 30-, and 60-minute \ac{PH}. LES denotes the lowest $\textit{RMSE}_{N_{60}}$ achieved by each participant across all evaluated parameter subsets. Each boxplot corresponds to a different identified parameter subset. A star ($*$) indicates a statistically significant difference from the preceding error distribution.}
    \label{fig:rmse_all_day}
\end{figure*}

\begin{table*}[!t]
\centering
\caption{Prediction error distributions (mg/dL) on the {validation} dataset for the "all day" case under the {intervention-informed} and {forecasting} modes. $\textit{RMSE}_{N_{PH}}$ is reported for PHs of 15, 30, and 60 min, and the $\textit{RMSE}_{av}$ over the complete 60-min PH. Values are reported as median [IQR].}
\label{tab:rmse_all_day}
\renewcommand{\arraystretch}{1.2}
\resizebox{\textwidth}{!}{%
\begin{tabular}{lcccccccc}
\toprule
\multirow{2}{*}{\textbf{Parameter subset}} &
\multicolumn{4}{c}{\textbf{{Intervention-informed}}} &
\multicolumn{4}{c}{\textbf{{Forecasting}}} \\
\cmidrule(lr){2-5}
\cmidrule(lr){6-9}
& \textbf{$\textit{RMSE}_{N_{15}}$} & \textbf{$\textit{RMSE}_{N_{30}}$} & \textbf{$\textit{RMSE}_{N_{60}}$} & \textbf{$\textit{RMSE}_{av}$}
& \textbf{$\textit{RMSE}_{N_{15}}$} & \textbf{$\textit{RMSE}_{N_{30}}$} & \textbf{$\textit{RMSE}_{N_{60}}$} & \textbf{$\textit{RMSE}_{av}$} \\
\midrule

Population
& \begin{tabular}[c]{@{}c@{}}16.60\\[-1mm]\footnotesize[13.22, 19.91]\end{tabular}
& \begin{tabular}[c]{@{}c@{}}37.92\\[-1mm]\footnotesize[29.60, 45.98]\end{tabular}
& \begin{tabular}[c]{@{}c@{}}69.68\\[-1mm]\footnotesize[55.36, 83.83]\end{tabular}
& \begin{tabular}[c]{@{}c@{}}32.72\\[-1mm]\footnotesize[17.75, 52.51]\end{tabular}
& \begin{tabular}[c]{@{}c@{}}16.62\\[-1mm]\footnotesize[13.24, 19.93]\end{tabular}
& \begin{tabular}[c]{@{}c@{}}38.24\\[-1mm]\footnotesize[29.83, 46.46]\end{tabular}
& \begin{tabular}[c]{@{}c@{}}70.78\\[-1mm]\footnotesize[56.58, 85.61]\end{tabular}
& \begin{tabular}[c]{@{}c@{}}33.52\\[-1mm]\footnotesize[18.38, 53.35]\end{tabular}
\\

$V_I, k_e, V_G,\tau_S,SI_2,\tau_D$
& \begin{tabular}[c]{@{}c@{}}6.60\\[-1mm]\footnotesize[5.66, 7.63]\end{tabular}
& \begin{tabular}[c]{@{}c@{}}16.75\\[-1mm]\footnotesize[14.10, 19.12]\end{tabular}
& \begin{tabular}[c]{@{}c@{}}30.95\\[-1mm]\footnotesize[26.02, 36.31]\end{tabular}
& \begin{tabular}[c]{@{}c@{}}11.48\\[-1mm]\footnotesize[6.47, 20.15]\end{tabular}
& \begin{tabular}[c]{@{}c@{}}6.59\\[-1mm]\footnotesize[5.65, 7.64]\end{tabular}
& \begin{tabular}[c]{@{}c@{}}16.76\\[-1mm]\footnotesize[14.08, 19.01]\end{tabular}
& \begin{tabular}[c]{@{}c@{}}31.31\\[-1mm]\footnotesize[26.20, 37.10]\end{tabular}
& \begin{tabular}[c]{@{}c@{}}11.44\\[-1mm]\footnotesize[6.44, 20.12]\end{tabular}
\\

$V_I, k_e, V_G,\tau_S,SI_2,\tau_D,k_{12}$
& \begin{tabular}[c]{@{}c@{}}6.55\\[-1mm]\footnotesize[5.60, 7.65]\end{tabular}
& \begin{tabular}[c]{@{}c@{}}16.55\\[-1mm]\footnotesize[14.13, 19.11]\end{tabular}
& \begin{tabular}[c]{@{}c@{}}30.68\\[-1mm]\footnotesize[25.96, 36.24]\end{tabular}
& \begin{tabular}[c]{@{}c@{}}11.41\\[-1mm]\footnotesize[6.44, 19.98]\end{tabular}
& \begin{tabular}[c]{@{}c@{}}6.54\\[-1mm]\footnotesize[5.60, 7.65]\end{tabular}
& \begin{tabular}[c]{@{}c@{}}16.60\\[-1mm]\footnotesize[14.06, 19.24]\end{tabular}
& \begin{tabular}[c]{@{}c@{}}30.74\\[-1mm]\footnotesize[26.01, 36.61]\end{tabular}
& \begin{tabular}[c]{@{}c@{}}11.41\\[-1mm]\footnotesize[6.43, 20.00]\end{tabular}
\\

$V_I, k_e, V_G,\tau_S,SI_2,\tau_D,k_{12},SI_1$
& \begin{tabular}[c]{@{}c@{}}6.47\\[-1mm]\footnotesize[5.42, 7.42]\end{tabular}
& \begin{tabular}[c]{@{}c@{}}16.49\\[-1mm]\footnotesize[13.85, 19.10]\end{tabular}
& \begin{tabular}[c]{@{}c@{}}30.88\\[-1mm]\footnotesize[25.91, 36.21]\end{tabular}
& \begin{tabular}[c]{@{}c@{}}11.38\\[-1mm]\footnotesize[6.39, 19.90]\end{tabular}
& \begin{tabular}[c]{@{}c@{}}6.47\\[-1mm]\footnotesize[5.42, 7.42]\end{tabular}
& \begin{tabular}[c]{@{}c@{}}16.60\\[-1mm]\footnotesize[13.84, 19.03]\end{tabular}
& \begin{tabular}[c]{@{}c@{}}30.76\\[-1mm]\footnotesize[25.91, 36.43]\end{tabular}
& \begin{tabular}[c]{@{}c@{}}11.37\\[-1mm]\footnotesize[6.39, 19.91]\end{tabular}
\\

\midrule
\textbf{LES}
& \begin{tabular}[c]{@{}c@{}}\textbf{6.37}\\[-1mm]\footnotesize\textbf{[5.38, 7.36]}\end{tabular}
& \begin{tabular}[c]{@{}c@{}}\textbf{16.32}\\[-1mm]\footnotesize\textbf{[13.71, 18.44]}\end{tabular}
& \begin{tabular}[c]{@{}c@{}}\textbf{30.13}\\[-1mm]\footnotesize\textbf{[24.91, 34.77]}\end{tabular}
& \begin{tabular}[c]{@{}c@{}}\textbf{11.08}\\[-1mm]\footnotesize\textbf{[6.27, 19.37]}\end{tabular}
& \begin{tabular}[c]{@{}c@{}}\textbf{6.37}\\[-1mm]\footnotesize\textbf{[5.37, 7.35]}\end{tabular}
& \begin{tabular}[c]{@{}c@{}}\textbf{16.25}\\[-1mm]\footnotesize\textbf{[13.74, 18.53]}\end{tabular}
& \begin{tabular}[c]{@{}c@{}}\textbf{30.19}\\[-1mm]\footnotesize\textbf{[25.14, 35.53]}\end{tabular}
& \begin{tabular}[c]{@{}c@{}}\textbf{11.08}\\[-1mm]\footnotesize\textbf{[6.25, 19.43]}\end{tabular}
\\

\bottomrule
\end{tabular}
}
\end{table*}

Performance is presented on the validation dataset in two cases: \say{all day} and \say{all events}. The \say{all day} case comprises predictions generated continuously throughout the entire recording period. In contrast, the \say{all events} case evaluates the model only around the main dynamic disturbances, considering 2-hour windows centered on \say{meal only}, \say{bolus injection}, and \say{postprandial response}.
Since the intervention-informed mode provides access to future inputs, allowing the model to anticipate upcoming disturbances, whereas the forecasting mode does not, the greatest differences between the two settings are expected to occur prior to these events. Consequently, the evaluation windows were centered on each disturbance to capture these differences.

\begin{figure*}[!t]  %
    \centering
    \includegraphics[trim=2mm 0mm 5mm 0mm, clip=true, width=\textwidth]{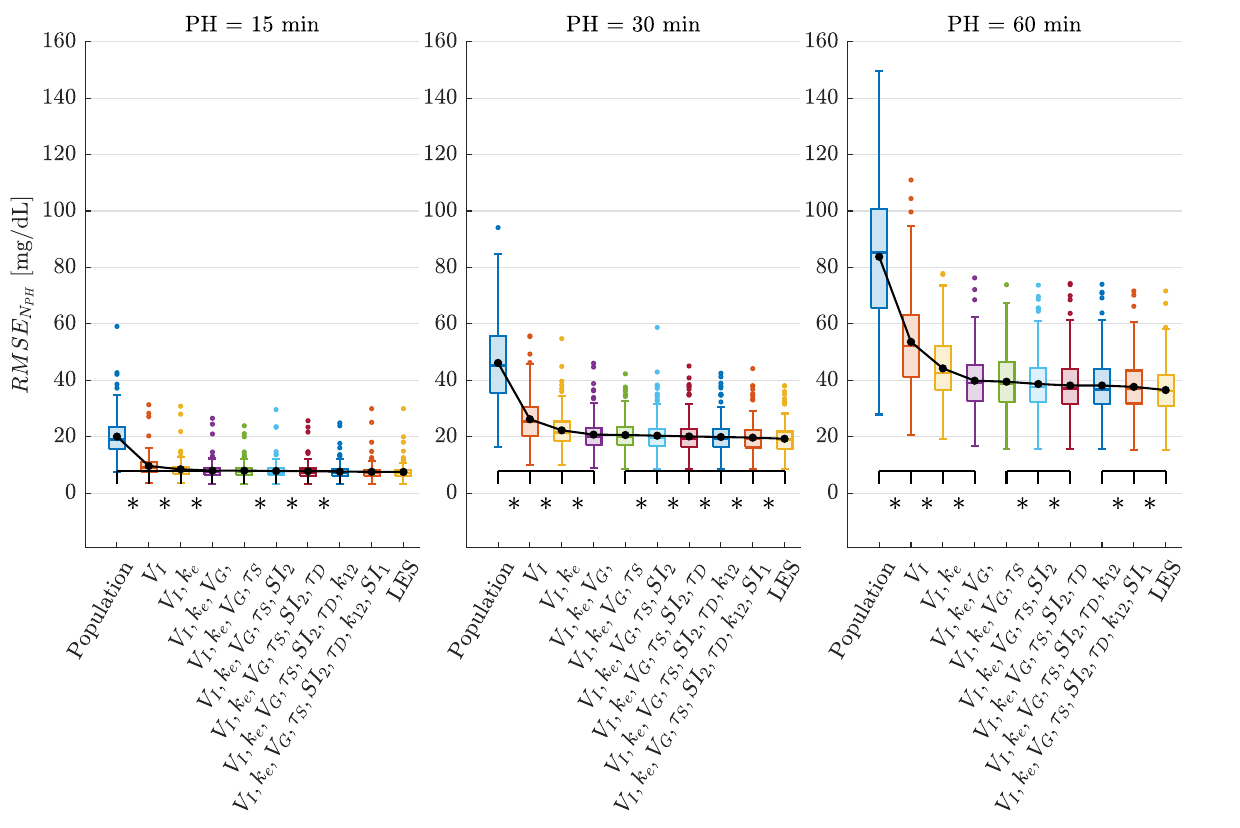}
    \caption{Distribution of validation $RMSE_{N_{PH}}$ values for all 192 participants in the "all events" at 15-, 30-, and 60-minute \ac{PH}. LES denotes the lowest $\textit{RMSE}_{N_{60}}$ achieved by each participant across all evaluated parameter subsets. Each boxplot corresponds to a different identified parameter subset. A star ($*$) indicates a statistically significant difference from the preceding error distribution.}
    \label{fig:rmse_pp}
\end{figure*}

Figures~\ref{fig:rmse_all_day} and~\ref{fig:rmse_pp} present the $\textit{RMSE}_{N_{PH}}$ at \ac{PH}s of 15, 30, and 60 minutes in the {intervention-informed} mode, which was used during parameter identification. As expected, prediction accuracy improves progressively as additional parameters are included in the identification strategy. \ac{LES} corresponds to the lowest $\textit{RMSE}_{N_{60}}$ achieved by each participant across all evaluated identified parameter subsets. Statistical differences between consecutive identification strategies were assessed using the paired Wilcoxon signed-rank test, as the prediction errors are paired and non-normally distributed. Significant differences with respect to the preceding strategy are indicated by stars ($*$) below each distribution.

Beyond the inclusion of $\tau_S$, the error distributions become nearly indistinguishable, suggesting that further parameter additions provide only marginal improvements at the population level. Nevertheless, the paired Wilcoxon test continues to identify statistically significant differences because it evaluates within-participant changes; consequently, even small but consistent reductions in individual prediction errors can result in statistical significance despite substantial overlap between the group-level distributions. A similar trend is observed for the \say{all events} case. 

\begin{table*}[!t]
\centering
\caption{Prediction error distributions (mg/dL) on the validation dataset for the "all events" case under the {intervention-informed} and {forecasting} modes. $\textit{RMSE}_{N_{PH}}$ is reported for PHs of 15, 30, and 60 min, and the $\textit{RMSE}_{av}$ over the complete 60-min PH. Values are reported as median [IQR].}
\label{tab:rmse_test_postprandial}
\renewcommand{\arraystretch}{1.2}
\resizebox{\textwidth}{!}{%
\begin{tabular}{lcccccccc}
\toprule
\multirow{2}{*}{\textbf{Parameter subset}} &
\multicolumn{4}{c}{\textbf{{Intervention-informed}}} &
\multicolumn{4}{c}{\textbf{{Forecasting}}} \\
\cmidrule(lr){2-5}
\cmidrule(lr){6-9}
& \textbf{$\textit{RMSE}_{N_{15}}$} & \textbf{$\textit{RMSE}_{N_{30}}$} & \textbf{$\textit{RMSE}_{N_{60}}$} & \textbf{$\textit{RMSE}_{av}$}
& \textbf{$\textit{RMSE}_{N_{15}}$} & \textbf{$\textit{RMSE}_{N_{30}}$} & \textbf{$\textit{RMSE}_{N_{60}}$} & \textbf{$\textit{RMSE}_{av}$} \\
\midrule

Population
& \begin{tabular}[c]{@{}c@{}}19.09\\[-1mm]\footnotesize[15.72, 23.34]\end{tabular}
& \begin{tabular}[c]{@{}c@{}}45.25\\[-1mm]\footnotesize[35.65, 55.57]\end{tabular}
& \begin{tabular}[c]{@{}c@{}}85.34\\[-1mm]\footnotesize[65.65, 100.71]\end{tabular}
& \begin{tabular}[c]{@{}c@{}}42.12\\[-1mm]\footnotesize[22.32, 66.05]\end{tabular}
& \begin{tabular}[c]{@{}c@{}}19.14\\[-1mm]\footnotesize[15.75, 23.52]\end{tabular}
& \begin{tabular}[c]{@{}c@{}}45.91\\[-1mm]\footnotesize[36.31, 56.44]\end{tabular}
& \begin{tabular}[c]{@{}c@{}}88.12\\[-1mm]\footnotesize[68.22, 102.69]\end{tabular}
& \begin{tabular}[c]{@{}c@{}}44.44\\[-1mm]\footnotesize[24.69, 67.43]\end{tabular}
\\

$V_I, k_e, V_G,\tau_S,SI_2,\tau_D$
& \begin{tabular}[c]{@{}c@{}}7.39\\[-1mm]\footnotesize[6.27, 8.78]\end{tabular}
& \begin{tabular}[c]{@{}c@{}}19.51\\[-1mm]\footnotesize[16.76, 22.77]\end{tabular}
& \begin{tabular}[c]{@{}c@{}}37.61\\[-1mm]\footnotesize[32.38, 44.39]\end{tabular}
& \begin{tabular}[c]{@{}c@{}}15.20\\[-1mm]\footnotesize[8.69, 25.96]\end{tabular}
& \begin{tabular}[c]{@{}c@{}}7.38\\[-1mm]\footnotesize[6.27, 8.79]\end{tabular}
& \begin{tabular}[c]{@{}c@{}}19.54\\[-1mm]\footnotesize[16.62, 22.95]\end{tabular}
& \begin{tabular}[c]{@{}c@{}}38.30\\[-1mm]\footnotesize[32.44, 45.03]\end{tabular}
& \begin{tabular}[c]{@{}c@{}}15.18\\[-1mm]\footnotesize[8.63, 26.11]\end{tabular}
\\

$V_I, k_e, V_G,\tau_S,SI_2,\tau_D,k_{12}$
& \begin{tabular}[c]{@{}c@{}}7.38\\[-1mm]\footnotesize[6.23, 8.74]\end{tabular}
& \begin{tabular}[c]{@{}c@{}}19.38\\[-1mm]\footnotesize[16.51, 22.86]\end{tabular}
& \begin{tabular}[c]{@{}c@{}}37.04\\[-1mm]\footnotesize[31.74, 44.02]\end{tabular}
& \begin{tabular}[c]{@{}c@{}}14.89\\[-1mm]\footnotesize[8.53, 25.47]\end{tabular}
& \begin{tabular}[c]{@{}c@{}}7.38\\[-1mm]\footnotesize[6.23, 8.72]\end{tabular}
& \begin{tabular}[c]{@{}c@{}}19.40\\[-1mm]\footnotesize[16.55, 22.90]\end{tabular}
& \begin{tabular}[c]{@{}c@{}}37.84\\[-1mm]\footnotesize[31.89, 44.56]\end{tabular}
& \begin{tabular}[c]{@{}c@{}}14.98\\[-1mm]\footnotesize[8.53, 25.70]\end{tabular}
\\

$V_I, k_e, V_G,\tau_S,SI_2,\tau_D,k_{12},SI_1$
& \begin{tabular}[c]{@{}c@{}}7.16\\[-1mm]\footnotesize[6.03, 8.56]\end{tabular}
& \begin{tabular}[c]{@{}c@{}}19.22\\[-1mm]\footnotesize[16.28, 22.57]\end{tabular}
& \begin{tabular}[c]{@{}c@{}}36.74\\[-1mm]\footnotesize[31.66, 43.90]\end{tabular}
& \begin{tabular}[c]{@{}c@{}}14.87\\[-1mm]\footnotesize[8.53, 25.42]\end{tabular}
& \begin{tabular}[c]{@{}c@{}}7.16\\[-1mm]\footnotesize[6.02, 8.55]\end{tabular}
& \begin{tabular}[c]{@{}c@{}}19.25\\[-1mm]\footnotesize[16.34, 22.76]\end{tabular}
& \begin{tabular}[c]{@{}c@{}}37.81\\[-1mm]\footnotesize[32.05, 44.82]\end{tabular}
& \begin{tabular}[c]{@{}c@{}}14.94\\[-1mm]\footnotesize[8.53, 25.62]\end{tabular}
\\

\midrule
\textbf{LES}
& \begin{tabular}[c]{@{}c@{}}\textbf{7.21}\\[-1mm]\footnotesize\textbf{[6.04, 8.19]}\end{tabular}
& \begin{tabular}[c]{@{}c@{}}\textbf{19.10}\\[-1mm]\footnotesize\textbf{[15.81, 21.83]}\end{tabular}
& \begin{tabular}[c]{@{}c@{}}\textbf{36.22}\\[-1mm]\footnotesize\textbf{[30.76, 41.85]}\end{tabular}
& \begin{tabular}[c]{@{}c@{}}\textbf{14.50}\\[-1mm]\footnotesize\textbf{[8.31, 24.56]}\end{tabular}
& \begin{tabular}[c]{@{}c@{}}\textbf{7.13}\\[-1mm]\footnotesize\textbf{[6.06, 8.21]}\end{tabular}
& \begin{tabular}[c]{@{}c@{}}\textbf{19.19}\\[-1mm]\footnotesize\textbf{[16.03, 22.04]}\end{tabular}
& \begin{tabular}[c]{@{}c@{}}\textbf{36.72}\\[-1mm]\footnotesize\textbf{[30.96, 43.17]}\end{tabular}
& \begin{tabular}[c]{@{}c@{}}\textbf{14.60}\\[-1mm]\footnotesize\textbf{[8.31, 24.73]}\end{tabular}
\\

\bottomrule
\end{tabular}
}
\end{table*}

\begin{figure}[!t]  %
    \centering
    \includegraphics[trim=7mm 0mm 9mm 0mm, clip=true, width=\linewidth]{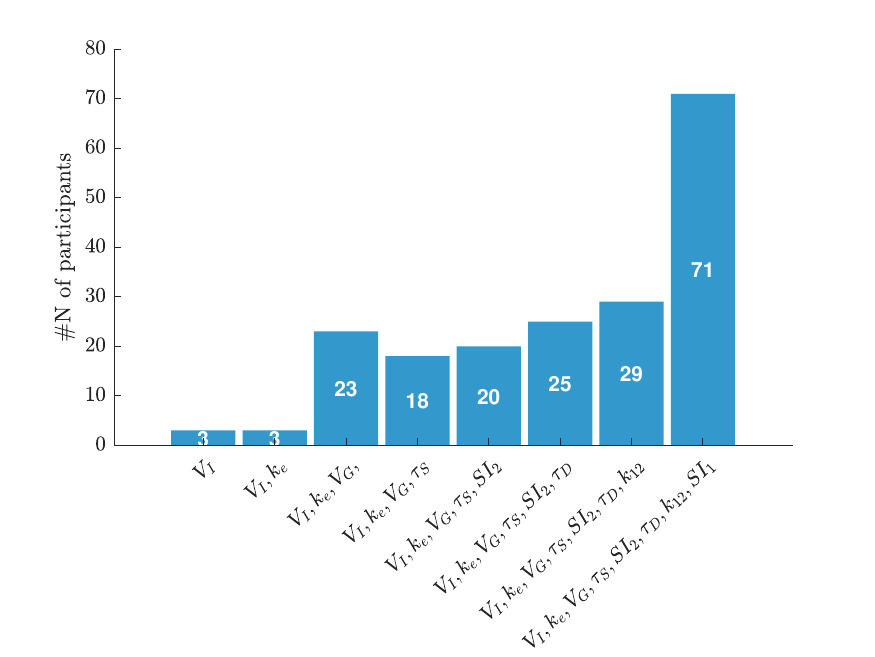}
    \caption{Number of participants achieving the lowest $RMSE_{N_{60}}$ with different identified parameter sets during "all day" case.}
    \label{fig:best_rmse_all_day}
\end{figure}

\begin{figure}[!t]  %
    \centering
    \includegraphics[trim=7mm 0mm 9mm 0mm, clip=true, width=\linewidth]{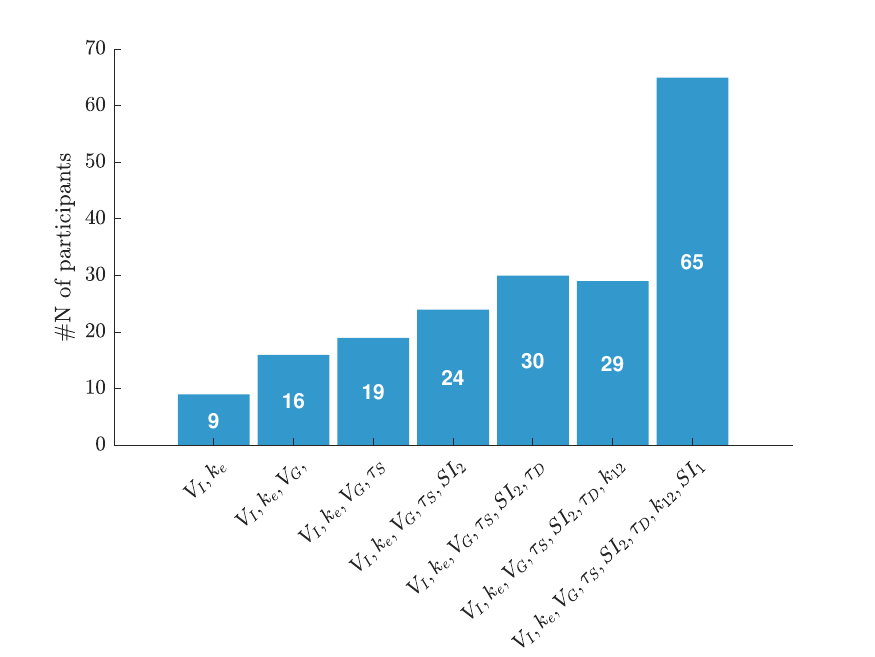}
    \caption{Number of participants achieving the lowest $RMSE_{N_{60}}$ with different identified parameter sets during "all events" case.}
    \label{fig:best_rmse_pp}
\end{figure}

Tables~\ref{tab:rmse_all_day} and~\ref{tab:rmse_test_postprandial} summarize the prediction performance across participants in the \say{all day} and \say{all events} cases, respectively, for both the interventional-informed and forecasting modes. Median $\textit{RMSE}_{N_{15}}$, $\textit{RMSE}_{N_{30}}$, and $\textit{RMSE}_{N_{60}}$ values are computed using \eqref{eq:rmse_fix_horizon}, and the $\textit{RMSE}_{av}$ as per \eqref{eq:RMSE_total}. Results are shown for the population parameters and the final three identified parameter sets, indicating comparable prediction error between the two modes. Figures~\ref{fig:best_rmse_all_day} and~\ref{fig:best_rmse_pp} show the number of participants for whom each parameter subset achieved the lowest $\textit{RMSE}_{N_{60}}$. In both scenarios, more participants achieve better predictions as parameters are progressively added, with $\tau_D$ standing out as important in the "all events" case, reflecting the importance of carbohydrate absorption dynamics.

\begin{figure*}[!t] %
    \centering
    \includegraphics[trim=12mm 0mm 10mm 0mm, clip=true, width=\textwidth]{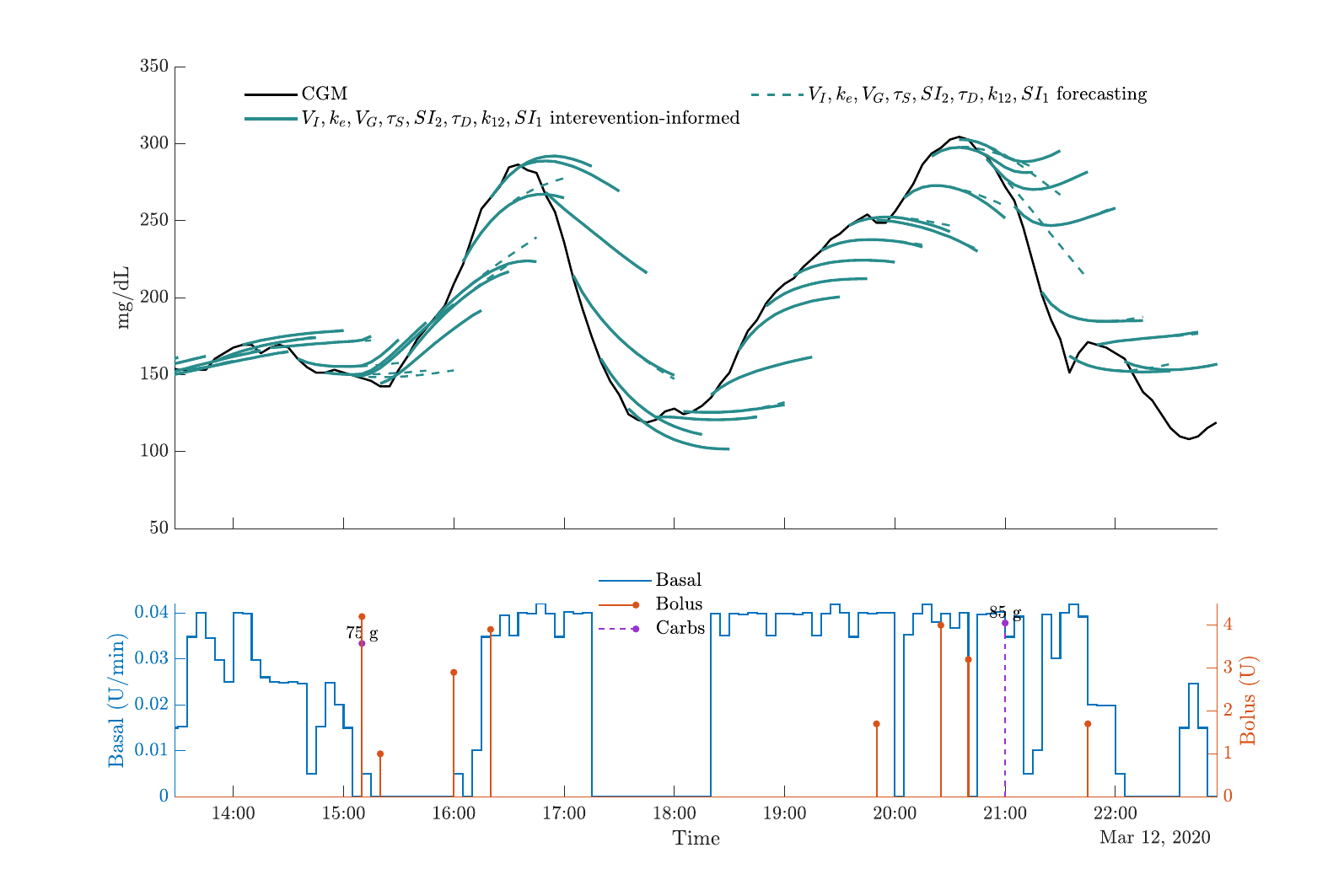}
     \caption{Representative 1-hour-ahead glucose predictions for a participant from the T1DEXI dataset. Solid teal lines correspond to the {informed} setting, in which future meal intake, insulin bolus administration, and the predefined basal insulin profile are assumed to be known over the prediction horizon. Dashed teal lines correspond to the {observational} setting, where future meal and bolus information are unavailable and the basal insulin infusion is assumed to remain constant at its initial value throughout the prediction horizon. The bottom panel shows insulin bolus doses, basal insulin infusion, and carbohydrate intake.}
    \label{fig:subj_6_predictions}
\end{figure*}

\begin{figure*}[!t]
    \centering
    \includegraphics[trim=12mm 0mm 10mm 0mm, clip=true, width=\textwidth]{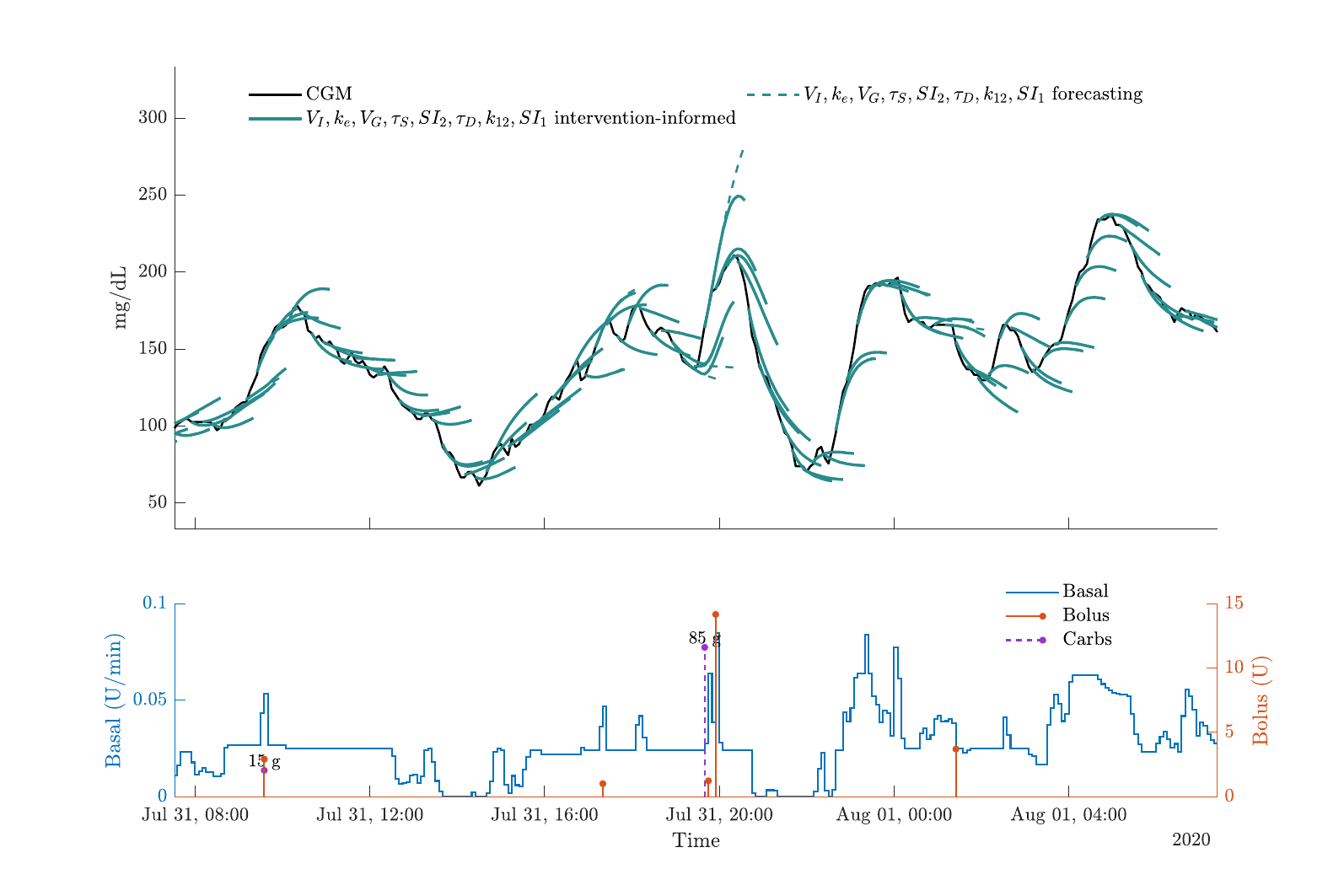}
    \caption{Representative 1-hour-ahead glucose predictions for a participant from the T1DEXI dataset. Solid teal lines correspond to the {informed} setting, in which future meal intake, insulin bolus administration, and the predefined basal insulin profile are assumed to be known over the prediction horizon. Dashed teal lines correspond to the {observational} setting, where future meal and bolus information are unavailable and the basal insulin infusion is assumed to remain constant at its initial value throughout the prediction horizon. The bottom panel shows insulin bolus doses, basal insulin infusion, and carbohydrate intake.}
    \label{fig:subj_11_predictions}
\end{figure*}

To complement these population-level results, Figures~\ref{fig:subj_6_predictions} and~\ref{fig:subj_11_predictions} present prediction examples for two participants over the validation day in both intervention-informed and forecasting modes. In each case, the identified parameter set corresponds to the one achieving the lowest $\mathrm{RMSE}_{N_{60}}$. Consistent with the quantitative results, prediction accuracy is remarkably similar in both modes. This is expected because the two approaches differ only when meal or bolus information falls within the \ac{PH}; otherwise, they receive the same inputs and therefore produce nearly identical predictions.

\begin{figure*}[!t] %
    \centering
    \includegraphics[trim=0mm 0mm 5mm 0mm, clip=true, width=\textwidth]{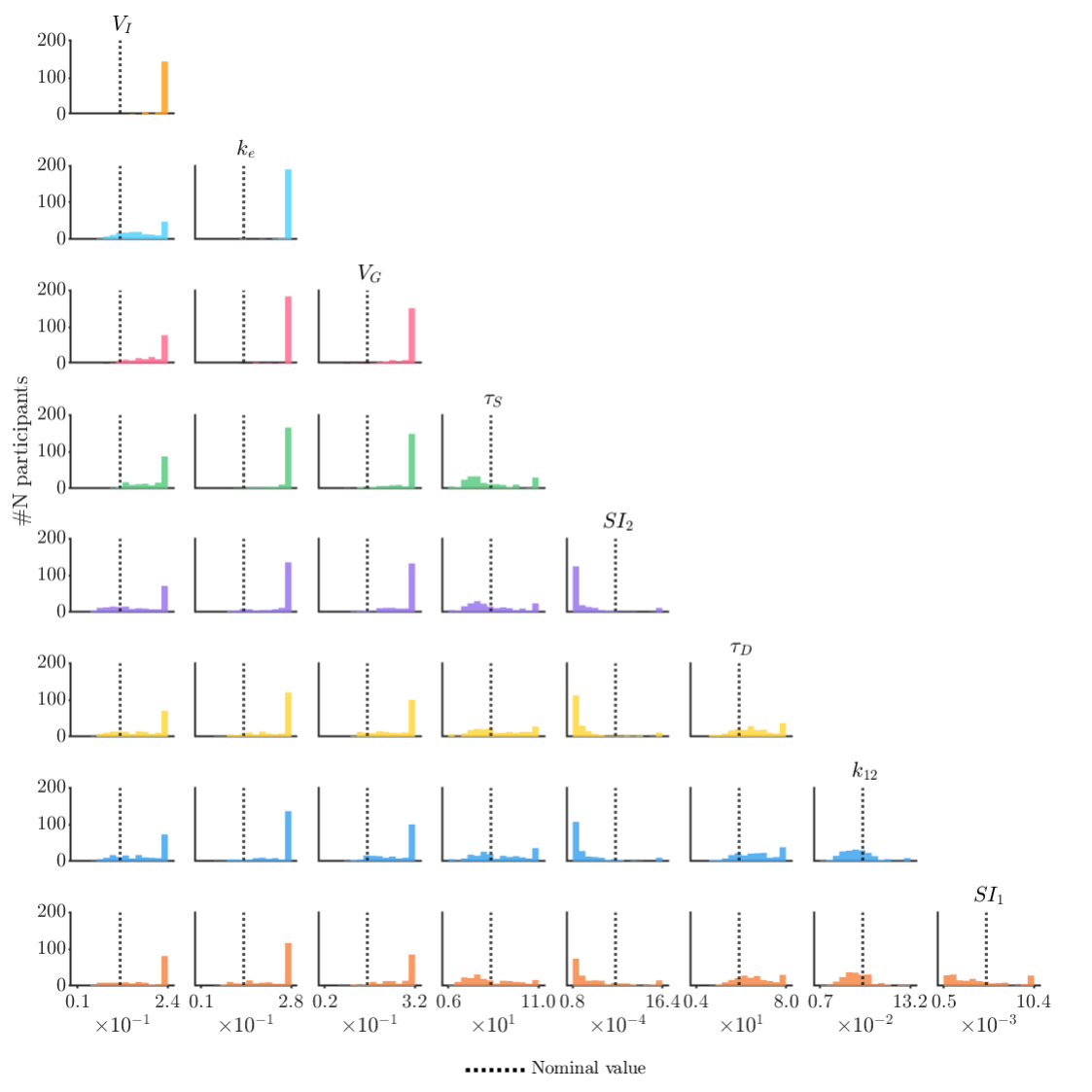}
    \caption{Distribution of the identified parameter values across the 192 participants. Each row corresponds to a different subset of identified parameters, ranging from the identification of only $V_I$ (top row) to the full set $\{V_I, k_e, V_G, \tau_S, SI_2, \tau_D, k_{12}, SI_1\}$ (bottom most row), with one additional parameter included at each successive step. The dotted vertical line denotes the nominal parameter value, and the x-axis limits correspond to the lower and upper bounds used during identification. Parameters' units are reported in Table~\ref{tab:HGM_summary}.}
    \label{fig:distribution_parameters}
\end{figure*}

Finally, Figure~\ref{fig:distribution_parameters} shows the distribution of the identified parameter values across participants. Each row corresponds to a different subset of identified parameters, ranging from the identification of only $V_I$ to the full set $\{V_I, k_e, V_G, \tau_S, SI_2, \tau_D, k_{12}, SI_1\}$, with one additional parameter included at each successive step. The upper and lower x-axis limits correspond to the parameter bounds used during identification.

%% file: Sections/Discussion.tex
\section{Discussion}
\label{sec:Discussion}

The global parameter rankings obtained for cohorts of  $\#N=10$, $\#N=20$, and $\#N=30$ participants from the \ac{T1DEXI} dataset consistently place $V_I$ and $k_e$ in the top two positions across all dynamic conditions. This can be explained since these parameters govern the insulin compartment and, consequently, regulate the amount of insulin available in the bloodstream. $V_G$ ranks closely behind, as it directly scales the $Q_1$ compartment, and gains prominence under basal insulin suppression. The relative importance of the remaining parameters varies across dynamic conditions; in particular, during postprandial responses, $\tau_S$ and $\tau_D$ become more influential. These observations underscore the importance of applying \ac{SA} across long data sequences encompassing diverse input conditions; otherwise, parameters that are critical for accurately capturing challenging dynamics, such as postprandial responses, may be overlooked.

It is worth noting that the sensitivity indices obtained here are smaller than those reported in~\cite{escorihuela2025parameters}. This discrepancy is likely due to the expansion of the sampling bounds from [0.5 1.5] to [0.1 2], as well as differences in the magnitude and scaling of the input data.
These observations highlight the importance of carefully selecting sampling ranges, which constitute a hyperparameter that the user must define. The choice should be guided both by the objectives of the \ac{SA} and, as in this case, by the requirements of the subsequent parameter identification procedure.

The stability of the Kendall's W statistic and its statistical significance as cohort size increases indicate that participants exhibit similar sensitivity patterns with respect to the model parameters. This consistency is expected: as shown, parameter ranking is primarily driven by the input data, and unless a participant's input profile deviates substantially from the rest of the cohort's distribution, the ranking remains largely stable. Consequently, while individual identification is still required to determine participants-specific parameter values, the subset of most influential parameters is consistent across participants. This finding is particularly relevant given the high computational cost of \ac{SA}, as identifying a common set of key parameters can substantially reduce the computation time required for model identification.

Although this study focused on parameter relevance and predictive performance, structural and practical identifiability were not explicitly investigated. Moreover, the Sobol \ac{SA} employed in this work assumes statistically independent input parameters, whereas some HGM parameters may exhibit correlations and compensatory effects. Consequently, parameters that rank highly in the \ac{SA} may still be strongly correlated with other parameters or be locally non-identifiable, making it difficult to uniquely estimate their values from the available data. Evidence of this behavior can be observed in Figure ~\ref{fig:distribution_parameters}, which shows the distribution of the identified parameter values across participants. In particular, the estimates of $V_I$, $k_e$, $V_G$, and $SI_2$ frequently converge to the upper or lower bounds of their admissible ranges, suggesting that the identification procedure may encounter local identifiability limitations or parameter compensation effects. Therefore, while \ac{SA} provides valuable information regarding parameter influence on model outputs, it should not be interpreted as a guarantee of parameter estimation. Incorporating structural and practical identifiability analysis into the proposed method represents an important direction for future work, as it could further refine the selection of personalized parameters while improving the robustness, interpretability, and clinical relevance of the identification process.

Beyond these methodological improvements, the proposed \ac{SA} can also be extended to other variants of the \ac{HGM}, as well as to other compartmental physiological models. Extending the \ac{HGM} to incorporate additional physiological processes would allow the framework to quantify the influence of the corresponding parameters and guide their prioritization during identification. The \ac{T1DEXI} dataset contains detailed information on physical activity, which is not explicitly represented in the baseline model. Incorporating exercise dynamics into the \ac{HGM} would enable the proposed methodology to determine the relative importance of exercise-related parameters under free-living conditions.

Regarding model performance, the reduction in $\textit{RMSE}_{av}$ and $\textit{RMSE}_{N_{PH}}$ at PHs of 15, 30 and 60 min when comparing population-based parameters with individually identified parameters from the \ac{LES} was approximately 60\% in both the {intervention-informed} and {forecasting} modes. This improvement highlights the limited predictive capability of population parameters and reinforces the need for individualized parameter identification.
Interestingly, no significant differences in performance metrics were observed between the {intervention-informed} and {forecasting} modes. For the \say{all day} case, this was expected because the two modes differ only for predictions preceding external disturbances, while most of the recording corresponds to basal conditions, resulting in similar average performance. For the \say{all events} case, reported meal announcements and bolus administrations do not always align perfectly with the subsequent glucose dynamics observed by the \ac{CGM}. Consequently, predictions generated in the {intervention-informed} mode do not necessarily match the measured glucose trajectory better than those obtained in the {forecasting} mode (see Figure~\ref{fig:subj_6_predictions}), resulting in only minor differences in overall prediction accuracy. It is also worth noting that prediction errors were consistently higher in the \say{all events} case than in the \say{all day} case. This behavior was expected, as meal intake and insulin administration induce rapid, highly nonlinear glucose-insulin dynamics that represent the most challenging conditions for physiological models to reproduce accurately.

These findings demonstrate that the proposed methodology provides consistent performance improvements across both prediction modes. Although the {interevention-informed} mode was evaluated over a 60-minute \ac{PH}, this choice was motivated by the intended applications of the personalized model rather than by the duration of the underlying physiological processes. Prediction horizons between 30 and 120 minutes are commonly adopted in model predictive control and other model-based decision support systems, where future interventions are assumed to be known or planned. We therefore selected a 60-minute horizon as a clinically relevant compromise for short-term decision support, while noting that the proposed parameter ranking methodology is not inherently limited to this horizon and can be readily extended to longer prediction windows.

In the forecasting mode, our approach compares favorably with existing identification strategies for glucose prediction in \ac{T1D} \cite{kushner2020,mosquera2022incorporating,zhu2023,prendin2023,giancotti2024forecasting, roquemen2026,basile2025}. Nevertheless, the comparison is only indicative, since the methods were evaluated on different datasets and under different experimental conditions. Despite substantial progress in personalized glucose forecasting, improvements in prediction accuracy have become progressively smaller with many recent approaches achieving comparable performance on free-living datasets. This pattern suggests an emerging "practical ceiling" that may reflect intrinsic limits imposed by delayed sensing (interstitial CGM), subcutaneous insulin pharmacodynamics, unobserved disturbances (e.g., imperfect meal/exercise reporting), and inter-/intra-individual variability, rather than the model class alone.

Finally, we emphasize that the use of $\textit{RMSE}_{av}$ and $\textit{RMSE}_{N_{PH}}$, was motivated to evaluate the quality of the identified models against available clinical observations. Specifically, they provide quantitative measures of how the proposed parameter ranking strategy influences the identification of the \ac{HGM} and its ability to reproduce observed glucose dynamics under clinically relevant conditions. Nevertheless, the appropriate validation metric, parameter fitting approach, and glucose forecasting method depend on the intended application of the personalized model.  While forecasting accuracy is suited for assessing predictive performance, it should not be regarded as a complete validation of a physiological digital twin. Applications involving counterfactual analysis, therapy planning, or decision support may require additional evaluation criteria, such as the physiological plausibility of the identified parameters and the consistency of model predictions under hypothetical interventions.

%% file: Sections/Conclusion.tex
\section{Conclusion}
\label{sec:conclusion}

A clearer understanding of how model parameters shape glucose-insulin dynamics is essential as \ac{DT} become more prominent in diabetes research. While these systems promise individualized metabolic simulations, their effectiveness relies on identifying and personalizing the parameters that have the greatest influence on model predictions under realistic and variable conditions.

This study quantified both the magnitude and the temporal relevance of the \ac{HGM} parameters across individuals to assess whether a universal parameter ranking is attainable or whether personalized parameter subsets are required when full input-output dynamics are considered. In light of this, we derived a global parameter ranking for the \ac{HGM} that accounts for inter‑subject variability while incorporating diverse real-life dynamic conditions, eliminating the need to perform computationally intensive global \ac{SA} for every new individual and substantially simplifying the parameter identification process.

Using this ranking, we demonstrated that accurate identification can be achieved only with a small subset of the most influential parameters, while additional parameters provide only marginal improvements in prediction accuracy. Furthermore, we showed that identified models achieve nearly identical glucose prediction performance in both the intervention-informed mode, where future meal and insulin information are available, and the forecasting mode, where such information is unavailable. This result indicates that the proposed identification strategy remains robust even when future meal and insulin information are unavailable, supporting its applicability in real-world forecasting and decision-support systems.

Importantly, while this study focused on the \ac{HGM}, the proposed \ac{SA} is generally applicable to any compartmental physiological model, providing a systematic approach to prioritize parameters for efficient and robust identification.

%% file: Sections/declarations.tex
\section*{Acknowledgment}
This manuscript is based on research using data from the T1DEXI adult cohort that has been made available through Vivli, Inc. Vivli has not contributed to or approved, and is not in any way responsible for, the contents of this publication.

\section*{Ethics statement}

This study did not involve human participants or animals. Therefore, ethical approval and informed consent were not required.

\section*{Funding statement}

The authors received no external funding for this study.
\section*{CRediT authorship contribution statement}

\textbf{C. Escorihuela-Altaba}: Writing - review \& editing, Writing - original draft, validation, investigation, formal analysis, conceptualization. \textbf{V. Naik \& E. Manzoni}: Supervision, writing - review \& editing, formal analysis, conceptualization. \textbf{J. Garcia-Tirado}: Supervision, writing - review \& editing, formal analysis, funding acquisition, conceptualization.

\section*{Declaration of competing interest}

The authors declare that they have no known competing financial interests or personal relationships that could have appeared to influence the work reported in this paper.

%% file: Sections/Appendix.tex
\renewcommand{\thetable}{A\arabic{table}}

\section*{List of acronyms}
AID-Automated Insulin Delivery, CGM-Continuous Glucose Monitoring, CHO-Carbohydrate, DT-Digital Twin, HGM-Hovorka Glucoregulatory Model, LES-Lowest Error Strategy, LHC-Latin Hypercube, MDI-Multiple Daily Injections, PH-Prediction Horizon, RMSE-Root Mean Square Error, SA-Sensitivity Analysis, SAP-Sensor-Augmented Pump, T1D-Type 1 Diabetes, T1DEXI-Type 1 Diabetes Exercise Initiative.

\section*{Appendix}
\section[Hovorka glucoregulatory model]{\raggedright Hovorka glucoregulatory model} \label{sec:appendix Hovorka}
\setcounter{table}{0}
The model used in this work is the \ac{HGM} \cite{hovorka-populational-parameters, Wilinska_Hovorka_oralmodel} which consists of the subcutaneous insulin absorption submodel, the glucose-regulatory system submodel, and the \ac{CHO} absorption submodel summarized as follows:
\begin{subequations}
    \label{eq:equations_model}
\begin{align} 
\frac{dQ_1(t)}{dt} &= U_G(t) - x_1(t)Q_1(t) - F_{c01}(t) - F_R(t) \notag \\\label{eq:H_1}
&\quad + k_{12}Q_2(t) + EGP(t) \\ 
\frac{dQ_2(t)}{dt} &= x_1(t)Q_1(t) - [k_{12} + x_2(t)]Q_2(t) \\
\frac{dx_1(t)}{dt} &= -k_{a1}x_1(t) + k_{b1}I(t) \\
\frac{dx_2(t)}{dt} &= -k_{a2}x_2(t) + k_{b2}I(t) \\
\frac{dx_3(t)}{dt} &= -k_{a3}x_3(t) + k_{b3}I(t) \\
\frac{dS_1(t)}{dt} &= u(t) - \frac{S_1(t)}{\tau_S} \\
\frac{dS_2(t)}{dt} &= \frac{S_1(t)}{\tau_S} - \frac{S_2(t)}{\tau_S} \\
\frac{dI(t)}{dt} &= \frac{S_2(t)}{\tau_S V_I} - k_e I(t) \\
\frac{dD_1(t)}{dt} &= \frac{1000 \cdot AG}{M_{wg}} \, d(t) - \frac{D_1(t)}{\tau_D} \\
\frac{dD_2(t)}{dt} &= \frac{D_1(t)}{\tau_D} - U_G(t) 
\end{align}
\end{subequations}
The corresponding model output is:
\begin{equation}
{G}(t)=\frac{Q_1(t)}{V_G}
\end{equation}

\section[Tuning of the Extended Kalman Filter]{\raggedright Tuning of the Extended Kalman Filter} \label{sec:appendix EKF}
The \ac{EKF} was configured with an initial state covariance matrix $P_0 = 0.001 \cdot I_{10}$,
where $I_{10}$ denotes the $10 \times 10$ identity matrix. The process noise covariance matrix was set as $Q = \mathrm{diag}(10,\,10,\,0.1,\,0.1,\,0.01,\,0.01,\,0.01,\,0.01,\,100,\,100)$,
where each diagonal element 
is selected proportional to the squared magnitude of the corresponding state variable under steady-state conditions, ensuring consistent scaling of process uncertainty across states. The measurement noise variance was set as $R = 0.01$,
thereby assigning a high level of confidence to the \ac{CGM} measurements during filtering.